\documentclass[12pt]{article}
\usepackage{amscd,amssymb,amsmath,latexsym,enumerate}
\usepackage[mathscr]{euscript}
\usepackage{epsfig}
\usepackage{fancybox}
\usepackage{verbatim}
\usepackage{multirow}
\usepackage{stackengine}

\usepackage{color}

\title{Index pairings in presence of symmetries \\
with applications to topological insulators}

\author{Julian Gro{\ss}mann, Hermann Schulz-Baldes
\\
\\
{\small Department Mathematik, Friedrich-Alexander-Universit\"at Erlangen-N\"urnberg, Germany}
}
\date{ }

\date{ }

\newtheorem{theo}{Theorem}
\newtheorem{defini}{Definition}
\newtheorem{proposi}{Proposition}
\newtheorem{lemma}{Lemma}

\newtheorem{remark}{Remark}

\newcommand{\BM}{{\mathbb B}}
\newcommand{\CM}{{\mathbb C}}
\newcommand{\NM}{{\mathbb N}}
\newcommand{\RM}{{\mathbb R}}
\newcommand{\SM}{{\mathbb S}}
\newcommand{\TM}{{\mathbb T}}
\newcommand{\ZM}{{\mathbb Z}}

\newcommand{\FM}{{\mathbb F}}

\newcommand{\Aa}{{\cal A}}
\newcommand{\Ee}{{\cal E}}

\newcommand{\Ff}{{\cal F}}
\newcommand{\Gg}{{\cal G}}

\newcommand{\Tr}{\mbox{\rm Tr}}

\newcommand{\Cc}{{\cal C}}

\newcommand{\Hh}{{\cal H}}
\newcommand{\cH}{{\cal H}}

\newcommand{\one}{{\bf 1}}

\newcommand{\Ind}{{\rm Ind}} 
 
\newcommand{\Ker}{{\rm Ker}} 
\newcommand{\Ran}{{\rm Ran}} 
 
\newcommand{\spec}{{\rm spec}} 
 
\newcommand{\Inv}{{\rm Inv}}

\newcommand{\Maj}{{\mbox{\rm\tiny Maj}}}

\newcommand{\SPH}{S_{\mbox{\rm\tiny ph}}}
\newcommand{\STR}{S_{\mbox{\rm\tiny tr}}}
\newcommand{\RCH}{R_{\mbox{\rm\tiny ch}}}

\usepackage[normalem]{ulem}
\usepackage{marginnote}

\begin{document}

\maketitle

\begin{abstract}
In a basic framework of a complex Hilbert space equipped with a complex conjugation and an involution, linear operators can be real, quaternionic, symmetric or anti-symmetric, and orthogonal projections can furthermore be Lagrangian. This paper investigates index  pairings of projections and unitaries submitted to such symmetries. Various scenarios emerge: Noether indices can take either arbitrary integer values or only even integer values or they can vanish and then possibly have secondary $\ZM_2$-invariants. These general results are applied to prove index theorems for the strong invariants of disordered topological insulators. The symmetries come from the Fermi projection ($K$-theoretic part of the pairing) and the Dirac operator ($K$-homological part of the pairing depending on the dimension of physical space).
\end{abstract}


\section{Introduction}
\label{sec-intro}

Topological insulators are free Fermion systems with the following characteristics: 

\vspace{.2cm}

\noindent $\bullet$ The Fermi level lies either in a spectral gap or a region of dynamical Anderson localization.

\vspace{.1cm}

\noindent $\bullet$ The system satisfies one or a combination of time-reversal, particle-hole or chiral symmetry. 

\vspace{.1cm}

\noindent $\bullet$ The Fermi projection is topologically distinguishable from a trivial projection. 

\vspace{.2cm}

\noindent All this terminology will be made mathematically precise in Section~\ref{sec-overview}. There are other traits and features that go along with the non-trivial topology (such as delocalized surface states), but these aspects will not be further discussed here and the reader is referred to an abundant literature, in particular the papers cited in the bibliography. The main object of this paper is to establish the following:

\vspace{.2cm}

\noindent $\bullet$ A systematic construction of index pairings for strong invariants of topological insulators.

\vspace{.2cm}

\noindent For the complex classes (systems with no symmetry or only a chiral symmetry) this was already achieved in \cite{PLB,PS}. Furthermore, in prior papers \cite{Sch,DS2} such index theorems were proved for all cases in dimension $d=2$ where the index pairing is well-known from the theory of the integer quantum Hall effect \cite{ASS,BES}. This two-dimensional case is now briefly described as a warm-up. Let us consider tight-binding models on the Hilbert space $\ell^2(\ZM^2)\otimes\CM^N$ with a finite dimensional fiber $\CM^N$ which takes into account spin and other internal degrees of freedom (such as sublattice or particle-hole). Then, given the Hamiltonian $H$ and a Fermi level $\mu$, one associates a Fermi projection $P=\chi(H\leq \mu)$ which has rapidly decreasing off-diagonal matrix elements due to localization estimates \cite{BES}. Let $F=(X_1+\imath X_2)/|X_1+\imath X_2|$ be the Dirac phase constructed from the two components of the position operator. Then $T=PFP+\one-P$ with $\one$ denoting the identity is a Fredholm operator. Its index is known to be equal to the Hall conductance \cite{BES}. Now let us suppose that the Hamiltonian has a time-reversal invariance $\STR^* \overline{H}\STR=H$ implemented by a real unitary $\STR$ squaring to $-\one$, namely the spin is odd. Due to $F^*=\overline{F}$, this implies that the Fredholm operator satisfies $\STR^* T^*\STR=\overline{T}$ which immediately implies that the index vanishes. This does not mean that there may not be non-trivial topology in the Fermi projection though. Indeed, it was shown in the ground-breaking paper of Kane and Mele \cite{KM} that time-reversal invariant insulators with odd spin and in dimension $d=2$ can have two distinguishable ground states, defining two topologically different phases. These two phases can actually be distinguished by a secondary invariant of the above Fredholm operator given by $\Ind_2(T)=\dim(\Ker(T))\;\mbox{mod}\,2\,\in\ZM_2$. This is a well-defined homotopy invariant due to the symmetry relation  $\STR^* T^*\STR=\overline{T}$, see \cite{Sch} and below, and is therefore indeed an adequate phase label. It is also possible to implement other symmetries of the Hamiltonian on the Fredholm operator $T$. For example, if $H$ has an odd particle-hole symmetry in dimension $d=2$, then the index of $T$ is always even, see \cite{DS2} and below. 

\vspace{.2cm}

The basic idea of this paper is now to consider the index pairings of the complex classes which are known for arbitrary even dimensions \cite{PLB} and arbitrary odd dimensions \cite{PS}, and to implement the symmetries of the Fermi projection in order to deduce $\ZM$-, $\ZM_2$- and $2\,\ZM$-valued index pairings. While in dimension $d=2$ this essentially leads to Fredholm operators lying in the classifying spaces of Atiyah and Singer \cite{AS} (as in the case of a time-reversal symmetric system discussed above), new types of $\ZM_2$- and $2\,\ZM$-valued index pairings emerge in other dimensions, and we were unable to understand them in terms of the classifying spaces (this remains a possibility though). The existence and well-definedness of these pairings is rooted in Kramers degeneracy arguments presented in  Sections~\ref{sec-symp2Z} and \ref{sec-sympZ2} which we believe to be new also on a level of linear algebra. Section~\ref{sec-indsym} presents and proves all these index pairings and constitutes the mathematical core of the paper. We deliberately chose to formulate these mathematical results using only basic functional analytic terminology, hoping that this renders the main results more accessible to a wider audience. However, it is possible to understand and apply these index pairings in the realm of non-commutative geometry \cite{Con} as the result of a pairing of $KR$-groups \cite{Kar,Sc} with $KR$-cycles \cite{Kas,GVF}. This is explained in Section~\ref{sec-NCG} which hopefully prepares the ground for further applications of the functional analytic results of Section~\ref{sec-indsym}. 


\vspace{.2cm}

The remainder of the paper, Section~\ref{sec-overview} on topological insulators, can be seen as an example where all types of index pairings with symmetries appear.  In connection with topological insulators a major advantage of the index approach is that it allows to deal with systems with broken translation invariance. Most prior works as \cite{KM,FKM,RSFL,Kit,ASV,DG,KZ} were restricted to periodic systems, except for \cite{HL,GP,FM,Tan} which do not develop an index theory approach though. The indices constructed in this work allow to distinguish topological ground states (often also called quantum phases) of disordered systems or of local perturbations of periodic systems, and thereby prove a much stronger stability result on these systems. Let us add that the same strategy to implement symmetries also works for the boundary index pairings \cite{KRS,EG,QHZ}, an issue that will be further developed elsewhere.

\vspace{.2cm}

Finally let us point the reader to the appendix about the Clifford group (essentially given by the generators of the Clifford algebra). A detailed understanding of symmetry operations in this group is significant for the understanding of the symmetries of the Dirac operator (defining the particular $KR$-cycle). These symmetries essentially arise from the dimension of physical space and therefore these constructions are possibly useful for other applications as well. The appendix  also presents a complementary perspective on the link between the classical matrix groups and the Clifford generators, differing from the existing literature \cite{Por}, and this is of interest in connection with the Cartan-Altland-Zirnbauer classes \cite{AZ}. 

\vspace{.2cm}

\noindent {\bf Acknowledgements:} This work greatly profited from discussions with many participants of the program on ``Topological Phases of Quantum Matter'' held at the ESI, Vienna, during the summer of 2014, so we are particularly thankful to Martin Zirnbauer for organizing this event. We thank Johannes Kellendonk for pointing us to the work of Van Daele \cite{VD}. This led to improvements in Sections~\ref{sec-Kgroups} and \ref{sec-KRgroups} in the final version of the manuscript. We also thank the DFG for partial financial support.

\vspace{.2cm}

\noindent {\bf Note added in proof:} After this work was accepted for publication, there appeared the preprints \cite{Kel} and \cite{BCR} on very related matters.

\section{Index pairings with symmetries}
\label{sec-indsym}

\subsection{Index pairings}
\label{sec-IndPair}

The main object of study in this paper are bounded Fredholm operators $T$ on a complex Hilbert space $\Hh$ of the type
\begin{equation}
\label{eq-BasicFred}
T\;=\;PFP\;+\;\one-P
\;,
\end{equation}
where $P=P^2=P^*$ is an orthogonal projection and $F$ is a unitary operator. Such Fredholm operators typically arise from index theorems obtained via pairings of $K$-groups with Fredholm modules ($K$-homological part) in two complementary ways \cite{Con}. In the first one, $F$ specifies an odd $K$-group element and $P$ is the Hardy projection onto the positive frequencies of the Dirac operator. In the second one, $P$ specifies an even $K$-group element and $F$ is the Dirac phase of a Dirac operator having a chiral symmetry (so that its Hardy projection is specified by $F$).  The Noether index of the Fredholm operator $T$ (note that Fritz Noether was the first to exhibit a Fredholm operator with a non-trivial index and also proved the first index theorem \cite{Noe}) is as usual defined by
$$
\Ind(T)\;=\;\dim(\Ker(T))-\dim(\Ker(T^*))
\;.
$$ 
In the following, the operators $P$ and $F$ are supposed to have symmetries connected with real and quaternionic structures on $\Hh$. These symmetries can impose $P$ to be real or quaternionic or Lagrangian, and $F$ to be real or quaternionic or symmetric or anti-symmetry, see Section~\ref{sec-PF} for details. One possibility is that these symmetries do not restrict the value of $\Ind(T)\in\ZM$, another one that it is always even, and yet another one that $\Ind(T)=0$. In this last case, it may be possible to define a secondary invariant using
$$
\Ind_2(T)
\;=\;
\dim(\Ker(T))\,\mbox{\rm mod}\;2\;\in\;\ZM_2
\;,
$$
whenever it is well-defined in the sense that it is a constant under norm-continuous homotopies of $P$ and $F$ respecting the symmetries imposed and the Fredholm property of $T$. Section~\ref{sec-PF} systematically studies what type of index is well-defined for a pairing with given symmetries.

\vspace{.2cm}

Now index pairing of projections with projections and unitaries with unitaries will be introduced. This is rather unnatural (and useless) within the framework of complex $K$-theory and $K$-homology. However, it becomes interesting when pairing $KR$-groups with $KR$-homology cycles. Given a pair of two projections $P$ and $E$, the operators $\one-2E$ and $\one-2P$ are both unitary and there are thus two index pairings deduced from \eqref{eq-BasicFred}:
\begin{equation}
\label{eq-ProjProj}
T\;=\;P(\one\,-\,2\,E)P\,+\,\one-P
\;,
\qquad
T'\;=\;E(\one\,-\,2\,P)E\,+\,\one-E
\;,
\end{equation}
provided both operators are Fredholm. The following elementary result re-establishes the symmetry under exchange of $P$ and $E$. 

\begin{proposi}
\label{prop-IndTT'} 
For Fredholm operators $T$ and $T'$ given in {\rm \eqref{eq-ProjProj}}, one has $\Ind(T)=\Ind(T')=0$ and $\Ind_2(T)=\Ind_2(T')$.
\end{proposi}

\noindent {\bf Proof.} 
As both $T$ and $T'$ are self-adjoint, their indices vanish. Next let us check that their kernels are of same dimension. If $v\in\Ker(T)$, then $v=Pv\in\Ran(P)$ and $PEPv=\frac{1}{2}\,v$ because $T=\one-2\,PEP$. Hence $w=Ev$ satisfies $EPE w=\frac{1}{2}\,w$ so that $w\in\Ker(T')$. By the same argument $u=Ew\in\Ker(T)$. But $u=EPv=\frac{1}{2}\,v$ so that all maps are invertible.
\hfill $\Box$

\vspace{.2cm}

Next let be given two unitary operators $U$ and $F$. They lead to two projections on the Hilbert space $\cH \oplus \cH$ by setting $P=\frac{1}{2}\binom{\one\;\;U}{U^*\;\;\one}$ and $E=\frac{1}{2}\binom{\one\;\;F}{F^*\;\;\one}$. Consequently, there are again two index pairings deduced from \eqref{eq-BasicFred}:
\begin{equation}
\label{eq-UniUni}
T\;=\;P\begin{pmatrix} F & 0 \\ 0 & F\end{pmatrix} P\,+\,\one-P
\;,
\qquad
T'\;=\;E\begin{pmatrix} U & 0 \\ 0 & U\end{pmatrix} E\,+\,\one-E
\;.
\end{equation}
There are other possibilities than choosing the unitaries block diagonal, but \eqref{eq-UniUni} is the only acceptable choice in connection with commuting symmetry operators. The following is again elementary. 

\begin{proposi}
\label{prop-IndT''T'''} 
For  Fredholm operators $T$ and $T'$ given in {\rm \eqref{eq-UniUni}}, one has $\Ind(T)=\Ind(T')=0$ and 
$$
\Ind_2(T)
\;=\;
\Ind_2(T')
\;=\;
\dim(\Ker(UF+FU))\,\mbox{\rm mod}\,2
\;.
$$
\end{proposi}

\noindent {\bf Proof.} 
Writing $P$ explicitly in \eqref{eq-UniUni}, one realizes that vectors in the kernel of $T$ are of the form $\binom{U}{\one}v$ with $v\in \Ker(UF+FU)$. This implies the equality and thus also $\Ind_2(T)=\Ind_2(T')$. Furthermore, the kernels of $T^*$ and $U^*F^*+F^*U^*$ are in bijection. Now if $v\in \Ker(UF+FU)$, then $w=F^*U^*v$ lies in the kernel of $U^*F^*+F^*U^*$. Thus $\Ind(T)=0$.
\hfill $\Box$

\vspace{.2cm}

Let us summarize all the possible pairings between two projections $P$ and $E$ and two unitaries $U$ and $F$ in a table:

\begin{equation}
\label{tab-IndPair0}
\begin{tabular}{|c||c|c|}
\hline
 & $F$ & $E$
\\
\hline
\hline
$P$ & $PFP+\one-P$ & $P(\one-2E)P+\one-P$ 
\\
\hline
$U$ & $UF+FU$  & $EUE+\one-E$
\\
\hline
\end{tabular}
\end{equation}
It will always be assumed that the appearing operators are Fredholm. The pairings on the diagonal are analyzed in Section~\ref{sec-PF}, those in the upper right and lower left can only be $\ZM_2$-valued due to Propositions~\ref{prop-IndTT'} and \ref{prop-IndT''T'''} and are studied in Sections~\ref{sec-PE} and \ref{sec-FU} respectively.

\subsection{Pairing projections and unitaries with symmetries}
\label{sec-PF}

First of all, the symmetries will be introduced. As there are several real and quaternionic structures on the Hilbert space $\Hh$ involved, we choose them to be expressed in terms of one fixed  complex conjugation $\Cc:\Hh\to\Hh$ which is an anti-linear isometric map squaring to the identity. Then the complex conjugate of a bounded operator $A\in\BM(\Hh)$ is defined by $\overline{A}=\Cc A\Cc$ and its transpose  $A^t=(\overline{A})^*$ as the adjoint of the complex conjugate. Furthermore $A$ is called real if $\overline{A}=A$. The symmetries are then implemented by symmetry operators in the following sense.

\begin{defini}
\label{def-SymmetryOp} 
A symmetry operator on a Hilbert space $\Hh$ with complex conjugation $\Cc$ is a real unitary $S=\overline{S}=(S^*)^{-1}\in\BM(\Hh)$ squaring either to the identity or minus the identity and having eigenspaces of same dimension. In these respective cases, it is called even or odd. 
\end{defini}

Note that the identity itself is also an even symmetry in this sense and that $S\Cc$ is a real or quaternionic structure for an even or odd symmetry operator $S$. It will be convenient (even though somewhat unconventional) to use the following terminology. 

\begin{defini}
\label{def-Symmetries} 
An operator $A\in\BM(\Hh)$ is called even real or odd real if $S^*\overline{A}S=A$ for some even or odd symmetry $S$ respectively. An operator $A\in\BM(\Hh)$ is called even symmetric or odd symmetric if $S^*A^tS=A$ with an even or odd symmetry $S$ respectively.  An orthogonal projection $P$ is called even (symplectic) Lagrangian or odd (symplectic) Lagrangian if $S^*\overline{P}S=\one-P$ with an even or odd symmetry $S$ respectively. 
\end{defini}

Let us add a few comments. An odd real operator can rightfully be called quaternionic. For self-adjoint operators the notions of symmetry and reality coincide. An operator $A$ is odd symmetric w.r.t. $S$ if and only if $SA=-(SA)^t$ is antisymmetric. The range of the Lagrangian projections is half dimensional and is also called maximally isotropic. In the literature the term Lagrangian is usually only used for the real odd Lagrangian case. In connection with topological insulators, the Lagrangian nature of $P$ is also called an even or odd particle hole symmetry. Some further comments on the Lagrangian structure follow in Section~\ref{sec-symp2Z} below.

\vspace{.2cm}

As already indicated above, there will $\ZM$-index,  $2\,\ZM$-index and $\ZM_2$-index theorems. Moreover, it will be explained in the subsections below how to further distinguish these index theorems into real, quaternionic, odd Lagrangian and even Lagrangian types (denoted by the letters R, Q, S and O respectively, where the O alludes to the complex orthogonal groups that is connected to even Lagrangian projections, see Appendix~\ref{sec-CGLieG}). This nomenclature is chosen according to the symmetry relation of the projection involved in the pairing. The following result specifies which type of index is well-defined for given symmetries.

\begin{theo}
\label{theo-indexlist}
Let $T=PFP+\one-P$ be a Fredholm operator constructed from a projection $P$ and a unitary $F$,  and let $S$, $\Sigma$ and $\widehat{S}$ be three commuting symmetries. In the respective cases, the index pairing is of the type indicated:
\begin{center}
\begin{tabular}{|c|c c||c|c|c|c|}
\hline
& & & $d=2$ & $d=4$ & $d=6$ & $d=8$
\\
\hline
\multirow{2}{*}{$j$} & $[S,F]=0$ & $[\widehat{S},F]=0$ & $\Sigma^*F^t\Sigma=F$ & $\Sigma^* \overline{F}\Sigma=F$ & $\Sigma^*F^t\Sigma=F$ & $\Sigma^*\overline{F}\Sigma=F$
\\
 & $[\Sigma,P]=0$ &  & $\Sigma^2=\one$  & $\Sigma^2=-\one$ & $\Sigma^2=-\one$  &  $\Sigma^2=\one$
\\
\hline
\hline
$0$ & $S^*\overline{P}S=P$ & $S^2=\one$ & $0$ & {\rm R}-$2\,\ZM$ & {\rm R}-$\ZM_2$  & $\ZM$
\\
\hline
\multirow{2}{*}{$1$} & $S^*\overline{P}S=P$ & $S^2=\one$
& \multirow{2}{*}{$0$} & \multirow{2}{*}{$0$} & \multirow{2}{*}{$0$}  &  \multirow{2}{*}{{\rm O}-$\ZM_2$}
\\
& $\widehat{S}^*\overline{P}\widehat{S}=\one-P$ & $\widehat{S}^2=\one$
& &  &   &
\\
\hline
$2$ & $S^*\overline{P}S=\one-P$ & $S^2=\one$ & $\ZM$ & $0$ & {\rm O}-$2\,\ZM$ & {\rm O}-$\ZM_2$
\\
\hline
\multirow{2}{*}{$3$} & $S^*\overline{P}S=\one-P$ & $S^2=\one$
& \multirow{2}{*}{{\rm Q}-$\ZM_2$}  & \multirow{2}{*}{$0$} & \multirow{2}{*}{$0$} & \multirow{2}{*}{$0$}
\\
& $\widehat{S}^*\overline{P}\widehat{S}=P$ & $\widehat{S}^2=-\one$
& &  &  &
\\
\hline
$4$ & $S^*\overline{P}S=P$ & $S^2=-\one$  & {\rm Q}-$\ZM_2$ & $\ZM$ & $0$ & {\rm Q}-$2\,\ZM$
\\
\hline
\multirow{2}{*}{$5$} & $S^*\overline{P}S=P$ & $S^2=-\one$
& \multirow{2}{*}{$0$} &  \multirow{2}{*}{{\rm S}-$\ZM_2$} & \multirow{2}{*}{$0$} & \multirow{2}{*}{$0$}
\\
& $\widehat{S}^*\overline{P}\widehat{S}=\one-P$ & $\widehat{S}^2=-\one$
& &  &   &
\\
\hline
$6$ & $S^*\overline{P}S=\one-P$ & $S^2=-\one$  & {\rm S}-$2\,\ZM$ & {\rm S}-$\ZM_2$ & $\ZM$ & $0$
\\
\hline
\multirow{2}{*}{$7$} & $S^*\overline{P}S=\one-P$ & $S^2=-\one$
& \multirow{2}{*}{$0$} & \multirow{2}{*}{$0$} & \multirow{2}{*}{{\rm R}-$\ZM_2$} & \multirow{2}{*}{$0$}
\\
& $\widehat{S}^*\overline{P}\widehat{S}=P$ & $\widehat{S}^2=\one$
 & &  &   &
\\
\hline
\end{tabular}
\end{center}
For each entry $0$ one has $\Ind(T)=0$ and, moreover, there exists a path of index pairings with the given symmetries in an augmented Hilbert space which connects the pairing to a pairing with trivial $\ZM_2$-index.
\end{theo}

In the table, the labels $j$ and $d$ are introduced for pure convenience for the moment being, but it will become apparent later on that they have the following interpretations: $j$ labels the group $KR_j$ of projections having the symmetries listed in the table, and $d$ labels the symmtery class of the $KR$ cycle (only the $4$ even ones are listed here, the odd ones are dealt with later on), see Section~\ref{sec-NCG} for details. Furthermore, in the theory of topological insulators, $j$ labels the Cartan-Altland-Zirnbauer classes and $d$ the dimension of physical space, see Section~\ref{sec-overview}. Topological insulators provide examples of operators with non-vanishing invariants. However, let us briefly sketch how such examples can be constructed from basic mathematical objects.


\vspace{.2cm}

\noindent{\bf Examples} (This extends \cite{Sch}.) Let $\Hh=\ell^2(\ZM)\otimes \CM^2=\ell^2(\ZM)\oplus\ell^2(\ZM)$ be equipped with the ``standard'' complex conjugation. Further let $\pi$ be the (Hardy) projection on $\ell^2(\ZM)$ with range $\ell^2(\NM)$ and $V$ the left shift on $\ell^2(\ZM)$. Then set $P=\binom{\pi\;0}{0\;\pi}$ and $F=\binom{V\;\;0}{0\;\;V^*}$.  These operators satisfy $P=\overline{P}$ and $F=\overline{F}$. If one introduces $S_{\pm}=\Sigma_{\pm}=\binom{0\;\pm\one}{\one\;\;0}$, then also $S_\pm^*\overline{P}S_\pm=P$ and $\Sigma_\pm^*F^t\Sigma_\pm=F$. Then $T=PFP+\one-P$ is a Fredholm operator with $\Ind(T)=0$ and $\Ind_2(T)=1$. The pairing is of the types $(j,d)=(0,2)$ and $(j,d)=(4,6)$  (only commuting symmetries are considered). The index inquires a different stability for the two cases. For $(j,d)=(0,2)$, all perturbations have a non-trivial $\ZM_2$-index. On the other hand, for $(j,d)=(4,6)$ it is possible to find a perturbation $T'=P'F'P'+\one-P'$ respecting the symmetries such that $\Ind(T')=0$ and $\Ind_2(T')=0$. 

\vspace{.1cm}

Now let us construct examples with $j=2,6$. Set $P=\binom{\pi\;\;\;\;\;0\;}{0\;\one-\pi}$. Then, with $S_\pm$ as above, $S_\pm^*\overline{P}S_\pm=\one-P$. Now, with $F$ and $\Sigma_\pm$ also as above, the  Fredholm operator $T=PFP+\one-P$ has $\Ind(T)=2$ and $\Ind_2(T)=0$. This is stable in the cases $d=2,6$ (strictly speaking, the for $(j,d)=(2,6)$ and $(j,d)=(6,2)$ the symmetries are anti-commuting, but this can be fixed as in the proof of Propostion~\ref{prop-FU}). It is a fun exercise left to the reader to construct non-trivial examples for all remaining cases in a similar manner. If $F$ is interpreted as fixing an element in the $KR$-groups of the C$^*$-algebra $C(\SM^1)$ equipped with the involutive isomorphism $(\tau f)(z)=f(\overline{z})$, then these examples provide index theorems detecting all the non-trivial $KR$-elements.
\hfill $\diamond$

\vspace{.2cm}

The next comments on the theorem concern the connection to the classifying spaces for Real $K$-theory of Atiyah and Singer \cite{AS}. These eight classifying spaces were introduced in \cite{AS} as sets of skew-adjoint Fredholm operators on a real Hilbert space having essential spectrum to both sides of the origin and anti-commuting with a representation of a real Clifford algebra. Here it is more convenient to work with a representation of these spaces on an infinite dimensional complex Hilbert space $\Hh$ equipped with complex conjugation $\Cc$ and a fixed odd symmetry operator $\Theta$ (such operators always exist). Let $\FM(\Hh)$ denote the Fredholm operators on $\Hh$ and $\FM_*(\Hh)$ the self-adjoint Fredholm operators having essential spectrum both in $\RM_<$ and $\RM_>$. Then it is relatively straightforward to find real-linear bijective maps identifying the spaces of \cite{AS} with the following sets:
\begin{align*}
&
\FM_1\;=\;\{T=-T^t\in\FM_*(\Hh)\}\;,
& & 
\FM_2\;=\;\{T=-T^t\in\FM(\Hh)\}\;,
\\
&
\FM_3\;=\;\{T=\Theta^*\overline{T}\Theta\in\FM_*(\Hh)\}\;,
& & 
\FM_4\;=\;\{T=\Theta^*\overline{T}\Theta\in\FM(\Hh)\}\;,
\\
&
\FM_5\;=\;\{T=-\Theta^*\overline{T}\Theta\in\FM_*(\Hh)\}\;,
& & 
\FM_6\;=\;\{T=T^t\in\FM(\Hh)\}\;,
\\
&
\FM_7\;=\;\{T=\overline{T}\in\FM_*(\Hh)\}\;,
& & 
\FM_8\;=\;\{T=\overline{T}\in\FM(\Hh)\}\;.
\end{align*}
It is proved in \cite{AS} that $\FM_1$ has the same homotopy type as the stabilized orthogonal group $\mbox{\rm O}$ and the homotopy groups are given by $\pi_{i-1}(\mbox{\rm O})=\pi_0(\FM_{i})$ and that
\begin{equation}
\label{tab-HomO}
\begin{tabular}{|c||c|c|c|c|c|c|c|c|}
\hline
$i$ & $1$ & $2$ & $3$ & $4$ & $5$ & $6$ & $7$ & $8$  \\
\hline
$\pi_{0}(\FM_i)$  &$\ZM_2$ & $\ZM_2$ & $0$ & $2\,\ZM$ & $0$ & $0$ & $0$ & $\ZM$ 
\\
\hline
\end{tabular}
\end{equation}
Furthermore, the connected components of $\FM_j$ are labelled by $\Ind$ and $\Ind_2$ defined above and this also explains the entry $2\,\ZM$ for $\FM_4$ in \eqref{tab-HomO} because kernel and cokernel of $T\in\FM_4$ clearly carry a quaternionic structure. Now it can readily be checked that the Fredholm operators $T$ in row $j=0$ of the table in Theorem~\ref{theo-indexlist} lie in the classifying space $\FM_6$, $\FM_4$, $\FM_2$ and $\FM_8$ respectively, while those in the row $j=4$ in $\FM_2$, $\FM_8$, $\FM_6$ and $\FM_4$ (one uses $\Theta=\Sigma S$ and verifies that one can use $\Theta=\one$ whenever $\Theta^2=\one$). Below in Sections~\ref{sec-indquat} and \ref{sec-indoddsym} short and independent proofs (not invoking \cite{AS}) are given that the index pairings are indeed well-defined in rows $j=0$ and $j=4$. What is not shown and investigated here is that the operator pairs $(P,F)$ and $(P',F')$ with same index can actually be homotopically deformed into each other without violating the symmetries and the Fredholm property. Such connectedness statements are known for the classifying spaces. The entries of the table carrying and S or an O as prefix require different arguments, which are presented in Sections~\ref{sec-symp2Z} and \ref{sec-sympZ2}. They provide new index pairings to our best knowledge, and we were not able to reduce them to the classifying spaces.

\vspace{.2cm}

The remainder of this section is mainly devoted to the proof of Theorem~\ref{theo-indexlist}. The basic well-known facts used here are that $\Ker(T)=\Ker(T^*T)$ and that $T$ is a Fredholm operator if and only if $0$ does not lie in the essential spectrum of $T^*T$. Therefore a basic property useful for assuring the existence of $2\,\ZM$ and $\ZM_2$-indices is a Kramers degeneracy (even multiplicity) of the (low lying) eigenvalues of $T^*T$. This results from linear algebra arguments which can become somewhat intricate though, see  Sections~\ref{sec-indquat} to \ref{sec-sympZ2} below. This Kramers degeneracy is sufficient for the proof of all $\ZM_2$-indices when combined with the following result.

\begin{proposi}
\label{prop-Ind0} 
For $T=PFP+\one-P$ one has $\Ind(T)=0$ in the following situations:

\vspace{.1cm}

\noindent {\rm (i)} $P$ is even or odd real and $F$ is even or odd symmetric.

\vspace{.1cm}

\noindent {\rm (ii)} $P$ is even or odd Lagrangian and $F$ is even or odd real.

\end{proposi}

\noindent {\bf Proof.} (i) The hypothesis imply that $(\Sigma S)^*T^t\Sigma S=T$ so that $\Ker(T^*)=\Cc\Sigma S\,\Ker(T)$.
(ii) \nolinebreak Here one checks $(\Sigma S)^*T^t\Sigma S=(\one-P)F^*(\one-P)+P$.  Combined with the following lemma this implies the claim.
\hfill $\Box$

\begin{lemma}
\label{lem-kernellink} For a unitary $F$ and a projection $P$, the map $F$ sends $\Ker(PFP+\one-P)$ bijectively to $\Ker((\one-P)F^*(\one-P)+P)$.
\end{lemma}

\noindent {\bf Proof.} \cite{DS2} 
Let $v\in\Ker(PFP+\one-P)$. Then $v\in\Ran(P)$. As $(\one-P)Fv=Fv-PFPv=Fv$ one deduces $Fv\in\Ran(\one-P)=\Ker(P)^\perp$.  Additionally, the calculation
$$
(\one-P) F^* (\one-P)Fv 
\; =\; 
(\one-P) F^* F v 
\;=\;
(\one- P) v 
\;=\; 
0
$$
shows that $F v \in \Ker((\one-P)F^*(\one-P)+P)$. Exchanging roles, this also shows that $F^*$ maps 
$\Ker((\one-P)F^*(\one-P)+P)$ into $\Ker(PFP+\one-P)$. 
\hfill $\Box$

\subsubsection{Quaternionic $2\,\ZM$-indices}
\label{sec-indquat}

In this section, the two $2\,\ZM$-index theorems in the rows $j=0$ and $j=4$ of the table of Theorem~\ref{theo-indexlist} are proved. In both cases, one uses the odd symmetry $\Theta=\Sigma S$ which then induces the odd reality relation $T=\Theta^*\overline{T}\Theta$, namely the matrix entries of $T$ are quaternions in the grading of $\Theta$. Hence $T$ is in the classifying space $\FM_4$ for which it is now shown (as in \cite{Sch}) that the index is even. This is based on the standard Kramers degeneracy argument presented below. It applies, in particular, to the kernel of $T$ and its cokernel $\Ker(T^*)=\Ker(TT^*)$, showing the R-$2\,\ZM$ and Q-$2\,\ZM$ entries in the table of Theorem~\ref{theo-indexlist}.

\begin{proposi}
\label{prop-posKramersQuat} 
Let $\Theta$ be an odd symmetry, and let $T=\Theta^*\overline{T}\Theta\in\BM(\Hh)$. Then
$$
T^*T\,v\;=\;\lambda\, v
$$
for some $\lambda\geq 0$ implies that
$$
w\;=\;\Theta\,\overline{v}
$$
is linearly independent of $v$ and also satisfies
$$
T^*T\,w\;=\;\lambda\, w
\;.
$$
Upon iteration one deduces that each eigenvalue of $T^*T$ has even degeneracy. 
\end{proposi}

\noindent {\bf Proof.} One has
$$
T^*Tw
\;=\;
\Theta\,(\Theta^*T\Theta)\,(\Theta^*T\Theta)\,\overline{v}
\;=\;
\Theta\;\overline{T^*T\,v}
\;=\;
\lambda\,\Theta\,\overline{v}
\;.
$$
Suppose now that there is $\mu\in\CM$ such that $v=\mu\,w$. Then one would have
$$
v\;=\;
\mu \, \Theta\,\overline{v}
\;=\;
\mu \, \Theta\,\overline{\mu \, \Theta\,\overline{v}}
\;=\;
-|\mu|^2 \,v
\;,
$$
in contradiction with $v\not = 0$. Now span$\{v,w\}$ is an invariant subspace of $T$ and so is its orthogonal complement. One now goes on and restricts $T$ to this orthogonal complement and repeats the above argument.
\hfill $\Box$

\subsubsection{Odd symmetric $\ZM_2$-indices}
\label{sec-indoddsym}

In the cases of the R-$\ZM_2$ and Q-$\ZM_2$ entries in the table of Theorem~\ref{theo-indexlist} the Fredholm operator $T$ is odd symmetric $\Theta^* T^t\Theta=T$ with an odd symmetry operator given by $\Theta=\Sigma S$. In other words, $T\in\FM_2$. By Proposition~\ref{prop-Ind0} this implies $\Ind(T)=0$. That the parity of the dimension of the kernel (nullity) of an odd symmetric Fredholm operator is a homotopy invariant was already proved in \cite{Sch}, and also follows from the results of \cite{AS}. Here a simple new proof is provided by showing that the all positive eigenvalues of $T^*T$ have an even multiplicity which implies that under homotopies the parity of the nullity is conserved.  This proves the entries R-$\ZM_2$ and Q-$\ZM_2$ of the table in Theorem~\ref{theo-indexlist}. Let us point out the following result does {\it not} apply for $\lambda=0$. If it applied, all $\ZM_2$-indices would be trivial.

\begin{proposi}
\label{prop-posKramers0} 
Let $\Theta$ be an odd symmetry and let $\Theta^*T^t\Theta=T\in\BM(\Hh)$ and $\lambda\not =0$. Then
$$
T^*T\,v\;=\;\lambda\, v
\;,
$$
implies that
$$
w\;=\;\Theta\,\overline{Tv}
\;,
$$
is linearly independent of $v$ and also satisfies
$$
T^*T\,w\;=\;\lambda\, w
\;.
$$
Upon iteration one deduces that each non-zero eigenvalue of $T^*T$ has even degeneracy. 
\end{proposi}

\noindent {\bf Proof.} One has
$$
T^*Tw
\;=\;
\Theta\,(\Theta^*T^*\Theta)\,(\Theta^*T\Theta)\,\overline{Tv}
\;=\;
\Theta\;\overline{T}\,\overline{T^*\,T\,v}
\;=\;
\lambda\,\Theta\,\overline{T}\,\overline{v}
\;=\;
\lambda\,w
\;.
$$
Suppose now that there is $\mu\in\CM$ such that $v=\mu\,w$. Then one would have
$$
v\;=\;
\mu \, \Theta\,\overline{T}\,\overline{v}
\;=\;
\mu \, \Theta\,\overline{T}\,\overline{\mu} \, \Theta\,T\,v
\;=\;
-|\mu|^2 \,T^*T\,v
\;=\;
-|\mu|^2\,\lambda\,v
\;,
$$
which is a contradiction to $v\not = 0$. Now span$\{v,w\}$ is an invariant subspace of $T^*T$ and so is its orthogonal complement. One now goes on and restricts $T^*T$ to this orthogonal complement and repeats the above argument.
\hfill $\Box$



\subsubsection{Symplectic $2\,\ZM$-indices}
\label{sec-symp2Z}

In this section the two commuting symmetries $S$ and $\Sigma$ satisfying $S^2=\eta\one$ and $\Sigma^2=-\eta\,\one$ for $\eta\in\{-1,+1\}$. Then $P$ is supposed to be Lagrangian w.r.t. $S$ and $F$ symmetric w.r.t. $\Sigma$, and both commuting with the other symmetry. Again $T=PFP+\one-P$ is supposed to be a Fredholm operator. This corresponds to the O-$2\,\ZM$ and S-$2\,\ZM$ entries in rows $j=2$ and $j=6$ of Theorem~\ref{theo-indexlist}. For the proofs below it will be convenient to work with frames instead of projections.

\begin{defini}
\label{def-SymplFrame} 
Let $S$ be a symmetry operator on $\Hh$. A (symplectic) Lagrangian frame w.r.t. $S$ is a linear map $\Phi:\Hh_0\to\Hh$ from some auxiliary Hilbert space $\Hh_0$ with complex conjugation satisfying 
$$
\Phi^*\Phi\;=\;\one_{\Hh_0}\;,
\qquad
\Phi\Phi^*\,+\,(S\overline{\Phi})(S\overline{\Phi})^*\;=\;\one_{\Hh}
\;.
$$
\end{defini}

Note that a symplectic frame $\Phi$ is a partial isometry. One immediately deduces $\Phi^*S\overline{\Phi}=0$ or equivalently $\Phi^tS\Phi=0$, namely column vectors in $\Phi$ and $S\overline{\Phi}$ are orthogonal in $\Hh$. Associated to $\Phi$ is a Lagrangian projection $P=\Phi\Phi^*$ in $\Hh$.
%
%
Next let us introduce
$$
T_0\;=\;\Phi^* T\Phi
\;=\;
\Phi^* F\Phi
\;\in\;\BM(\Hh_0)
\;.
$$
Clearly $T$ is Fredholm if and only if $T_0$ is Fredholm and $\Ind(T)=\Ind(T_0)$. Therefore the following result, applied to the kernel and cokernel of $T_0$, implies that  entries S-$2\,\ZM$ and O-$2\,\ZM$ in rows $j=2$ and $j=6$ of the table in Theorem~\ref{theo-indexlist} are well-defined.

\begin{proposi}
\label{prop-posKramers3} 
Let $S$, $\Sigma$, $P$, $F$, $\Phi$, $T$ and $T_0$ be as described in this subsection. Then
$$
T_0^*T_0\,v\;=\;\lambda\, v
\;,
\qquad
\lambda\geq 0
\;,
$$
implies that
$$
w\;=\;\overline{(S\overline{\Phi})^*\Sigma^*\,F\Phi\;v}
\;,
$$
is linearly independent of $v$ and also satisfies
$$
T_0^*T_0\,w\;=\;\lambda\, w
\;.
$$
Upon iteration one deduces that each eigenvalue of $T_0^*T_0$ has even degeneracy. 
\end{proposi}

\noindent {\bf Proof.}  First of all
\begin{align*}
T_0^*T_0
& = \;
\Phi^* F^*\Phi\Phi^* F\Phi
\;=\;
\one\;-\;
\Phi^* F^*(S\overline{\Phi})(S\overline{\Phi})^* F\Phi
\\
& =\;
\one\;-\;
\overline{\overline{\Phi}^* \Sigma^*F\Sigma(S{\Phi})(S{\Phi})^* \Sigma^*F^*\Sigma\overline{\Phi}}
\;=\;
\one\;-\;
\overline{(S\overline{\Phi})^* \Sigma^*F{\Phi}{\Phi}^* F^*\Sigma(S\overline{\Phi})}
\;.
\end{align*}
Hence
$$
T_0^*T_0w
\;=\;
\overline{(S\overline{\Phi})^* \Sigma^*F{\Phi}}\,(\one-\overline{{\Phi}^* \Sigma F^*(S\overline{\Phi})} \overline{(S\overline{\Phi})^* \Sigma^*F{\Phi}})\,\overline{v}
\;=\;
\overline{(S\overline{\Phi})^* \Sigma^*F{\Phi}}\;\overline{T_0^*T_0\,v}
\;=\;
\lambda \,w
\;.
$$
Now let $\mu\in\CM\setminus\{0\}$ be such that $v=\mu\,w$. Then
\begin{align*}
v\;
& =\;
\mu\,\overline{(S\overline{\Phi})^*\Sigma^*\,F\Phi\;v}
\;=\;
|\mu|^2\,\overline{(S\overline{\Phi})^*\Sigma^*\,F\Phi}(S\overline{\Phi})^*\,\Sigma^*F\Phi\;v
\\
& =\;
\eta\,|\mu|^2\,(S{\Phi})^*F^*\overline{\Phi}\,(S\overline{\Phi})^*F\Phi\;v
\; =\;
-\,|\mu|^2\,\eta^2\,\Phi^*F^*(S\overline{\Phi})\,(S\overline{\Phi})^*F\Phi\;v
\\
& =\;
-\,|\mu|^2\,(\one-T_0^*T_0)\,v
\;=\;
-\,|\mu|^2\,(1-\lambda)\,v
\;,
\end{align*}
which is impossible because $\lambda\in[0,1]$. 
\hfill $\Box$

\begin{remark}
\label{rem-antisym}
When $S^2=\one$, the above proof also goes through with minor modifications if the condition $\Sigma^*F^t\Sigma=F$ with $\Sigma^2=-\one$ is replaced by $F^t=-F$. Hence the Fredholm operator $T=PFP+\one-P$ for even real $P$ and anti-symmetric $F$ has even dimensional kernel and cokernel.
\end{remark}

\subsubsection{Symplectic $\ZM_2$-indices}
\label{sec-sympZ2}

In this section the two commuting symmetries $S$ and $\Sigma$ satisfying $S^2=\eta\one$ and $\Sigma^2=\eta\,\one$ for $\eta\in\{-1,+1\}$. Then $P$ is supposed to be Lagrangian w.r.t. $S$ and $F$ real w.r.t. $\Sigma$, and both commuting with the other symmetry. Again $T=PFP+\one-P$ is supposed to be a Fredholm operator. Hence the aim is to analyze the $\ZM_2$ entries in the rows $j=2$ and $j=6$ of Theorem~\ref{theo-indexlist}. In the following result, it is allowed to work with a second Lagrangian projection $Q$ also commuting with $\Sigma$. Choosing $P=Q$ and reasoning as in the first paragraph of Section~\ref{sec-indoddsym} then shows that the O-$\ZM_2$ and S-$\ZM_2$ entries in rows $j=2$ and $j=6$ of the table of Theorem~\ref{theo-indexlist} are well-defined. Again it is important to note that the argument does not apply to the eigenvalue $\lambda=0$.

\begin{proposi}
\label{prop-posKramers2} 
Let $S$, $\Sigma$, $P$, $Q$ and $F$ be as above, and further let $\Phi$ and $\Psi$ be Lagrangian frames for  $P$ and $Q$ respectively. Set
$$
T_0
\;=\;
\Psi^* F\Phi
\;.
$$
Then
$$
T_0^*T_0\,v\;=\;\lambda\, v
\;,
\qquad
\lambda>0
\;,
$$
implies that
$$
w\;=\;\overline{(S \overline{\Phi})^*\Sigma\,F^*\Psi\Psi^* F\Phi\;v}
\;,
$$
is linearly independent of $v$ and also satisfies
$$
T_0^*T_0\,w\;=\;\lambda\, w
\;.
$$
Upon iteration one deduces that each non-vanishing eigenvalue of $T_0^*T_0$ has even degeneracy. 
\end{proposi}

\noindent {\bf Proof.} Let us introduce the auxiliary operator 
$$
R_0\;=\;\Psi^*\,F\,S \,\Sigma \,\overline{\Phi}
\;.
$$
This allows to write $w=\overline{R_0^*T_0v}= R_0^t\overline{T}_0\,\overline{v}$. Furthermore, one has the identities
$$
T_0^t\,\overline{T}_0\;=\;\one-R^*_0\,R_0\;,
\qquad
 R_0^t\,\overline{T}_0\;=\;-\,\eta\,T_0^*\,R_0\;,
\qquad
T_0T_0^*\;=\;\one-R_0R_0^*
\;.
$$
Let us focus on the first two:
\begin{align*}
T_0^t\,\overline{T}_0
& =\;
\Phi^t\,\Sigma^*\,F^*\,\overline{\Psi}\,\Psi^t\,F\,\Sigma\,\overline{\Phi}
\; =\;
(S \overline{\Phi})^*\,\Sigma^*\,F^*\,(S \overline{\Psi})\,(S \overline{\Psi})^*\,F\,\Sigma\,(S \overline{\Phi})
\\
& =\;
(S \overline{\Phi})^*\,\Sigma^*\,F^*\,(\one-\Psi\Psi^*)\,F\,\Sigma^*(S \overline{\Phi})
\; =\;
\one\;-\;(S \overline{\Phi})^*\,\Sigma^*\,F^*\,\Psi\,\Psi^*\,F\,\Sigma\,(S \overline{\Phi})
\;=\;
\one-R_0^*R_0\;,
\\
R_0^t\,\overline{T}_0 &
=\;
(S  \Phi)^*\,F^t\,\overline{\Psi}\,\overline{\Psi}^*\,\overline{F}\,\overline{\Phi}\,
\;=\;
(S  \Phi)^*\,\Sigma^*\,F^*\,\overline{\Psi}\,\overline{\Psi}^*\,F\,\Sigma\,\overline{\Phi} 
\\
& =\;
\eta\,\Phi^*\,\Sigma^*\,F^*\,(S \overline{\Psi})\,(S \overline{\Psi})^*\,F\,\Sigma\,(S \overline{\Phi}) 
\;=\;
\eta\,\Phi^*\,\Sigma^*\,F^*\,(\one-{\Psi}\Psi^*)\,F\,\Sigma\,(S \overline{\Phi}) 
\;=\;
-\,\eta\,T_0^*\,R_0
\;.
\end{align*}
The third one follows in a similar manner. Using these identities, it can now be verified that $w$ is an eigenvector:
$$
T_0^*T_0\,w
\; =\,
\overline{T_0^t\overline{T}_0\,R_0^*\,T_0v}
\,=\,
\overline{(\one-R_0^*R_0)\,R_0^*\,T_0v}
\; =\,
\overline{R_0^*(\one-R_0R_0^*)\,T_0v}
\,=\,
\overline{R_0^*T_0T_0^*T_0v}
\,=\,
\lambda\;w
\;.
$$
Suppose now that there is $\mu\in\CM$ such that $v=\mu\,w$. Then one would have
\begin{align*}
T_0^*T_0\,v
& =\;
\mu\,T_0^*T_0 w
\;=\;
\mu\,T_0^*T_0 R_0^t\overline{T}_0\,\overline{v}
\;=\;
\eta\,|\mu|^2\,T_0^*T_0 \,(R_0^t\,\overline{T}_0)\,R_0^*\,T_0\,v
\\
& =\;
-\,\eta^2\,|\mu|^2\,T_0^*T_0 \,T_0^*\,R_0\,R_0^*\,T_0\,v
\;=\;
-|\mu|^2\,T_0^*T_0 \,T_0^*\,(\one-T_0T_0^*)\,T_0\,v
\;=\;
-|\mu|^2(\lambda^2-\lambda^3)v
\;,
\end{align*}
in contradiction to the positivity of $T_0^*T_0$ and $\|T_0\|\leq 1$ (so that $\lambda\leq 1$). Hence span$\{v,w\}$ is a two-dimensional invariant subspace of $T_0^*T_0$. Its orthogonal complement is also invariant. Restricting $T_0^*T_0$ to this orthogonal complement allows to repeat the argument.
\hfill $\Box$

\subsubsection{Vanishing index pairings with one symmetry each}
\label{sec-vanish}

The aim of this section is to justify the four entries $0$ in Theorem~\ref{theo-indexlist} with even $j$, namely for $(j,d)\in\{(0,2),\, (2,4),\,(4,6),\, (6,8)\}$. By Proposition~\ref{prop-Ind0} the Noether index vanishes in each of these cases so that only remains to show that the $\ZM_2$-index changes to the trivial value along a path of index pairings respecting the symmetries imposed. In all examples that we considered such a path could be constructed within the given Hilbert space (and we suspect this to be true in general), but here only the last claim of Theorem~\ref{theo-indexlist} will be proved, namely the path will be constructed in an augmented Hilbert space. Let us focus on the case $(j,d)=(0,2)$ since the others are dealt with in a similar manner. Set
$$
\widetilde{\Hh}\;=\;\Hh\oplus\CM^2
\;,
\qquad
\widetilde{\Sigma}\;=\;\Sigma\oplus\sigma_1
\;,
\qquad
\widetilde{S}\;=\;S\oplus\one_2
\;,
\qquad
\widetilde{\Cc}\;=\; \Cc\oplus\bar 
\;,
$$
where $\sigma_1=\binom{0\;1}{1\;0}$ and $\one_2=\binom{1\;0}{0\;1}$ and the overline denotes the complex conjugation in $\CM^2$. Furthermore with $r_\lambda= \binom{\cos(\lambda)\;-\sin(\lambda)}{\sin(\lambda)\;\;\cos(\lambda)}$ and $p=\binom{1\;0}{0\;0}$ let us set
$$
\widetilde{F}_\lambda\;=\;F\oplus r_\lambda
\;,
\qquad
\widetilde{P}\;=\;P\oplus p
\;,
\qquad
\widetilde{T}_\lambda\;=\;\widetilde{P}\widetilde{F}_\lambda\widetilde{P}\;
\,+\,\one-\widetilde{P}
\;.
$$
By construction $\widetilde{P}$ and $\widetilde{F}_\lambda$ satisfy all the symmetries of the index pairing $(j,d)=(0,2)$, and $\widetilde{T}_\lambda$ is a Fredholm operator for all $\lambda$. Now $\widetilde{T}_0=T\oplus\one_2$ so that, in particular, $\Ind_2(\widetilde{T}_0)=\Ind_2(T)$. On the other hand, at $\lambda=\frac{\pi}{2}$ the dimension of the kernel changes by $1$ so that $\Ind_2(\widetilde{T}_{\frac{\pi}{2}})=\Ind_2(T)+1$.

\subsubsection{Index pairings with three symmetries}
\label{sec-3sym}

Finally let us consider the cases in Theorem~\ref{theo-indexlist} where three symmetries $S$, $\widehat{S}$ and $\Sigma$ are involved. This corresponds to the rows with odd $j$. These rows are placed between two other rows which are specified by one of the symmetries for the projection $P$ each (cyclically, that is, the symmetries of $j=7$ are given by those of $j=6$ and $j=0$). As the symmetries $S$ and $\widehat{S}$ are supposed to commute, the Fredholm operators $T$ in rows with odd $j$ inherit all the properties from the neighboring even $j$'s. This implies that for each odd $j$ there is only one possibly non-vanishing entry in $\ZM_2$ which appears for $d$ such that for the neighboring $j$'s there are $\ZM$- and $\ZM_2$-indices. Indeed, if one of the neighboring $j$ has a $2\,\ZM$ entry, automatically the $\ZM_2$-index vanishes. This is relevant for $8$ of the $16$ cases, {\it e.g.} $(j,d)=(4,1),\,(6,1)$. The $0$ entries in the $4$ cases $(j,d)=(2,1),\,(4,3),\,(6,5),\,(8,7)$ follow from the arguments in Section~\ref{sec-vanish} because the homotopies constructed there merely modify $F$ and not $P$, so that they extend directly to the neighboring cases with an extra symmetry for $P$.

\subsection{Pairing projections with projections}
\label{sec-PE}

\begin{proposi}
\label{prop-PE} 
Let $P$ and $E$ be two projections such that $T=P(\one\,-\,2\,E)P\,+\,\one-P$ is a Fredholm operator. Then $\Ind_2(T)$ takes the following values:

\begin{center}
\begin{tabular}{|c|c c||c|c|c|c|}
\hline
& & & $d=1$ & $d=3$ & $d=5$ & $d=7$
\\
\hline
\multirow{2}{*}{$j$} & $[S,\Sigma]=0$ & $[S,E]=0$ & $\!\Sigma^*\overline{E}\Sigma=E\!$ & $\!\Sigma^* \overline{E}\Sigma=\one-E\!$ & $\!\Sigma^*\overline{E}\Sigma=E\!$ & $\!\Sigma^*\overline{E}\Sigma=\one-E\!$
\\
 & $[\Sigma,P]=0$ &  & $\Sigma^2=\one$  & $\Sigma^2=-\one$ & $\Sigma^2=-\one$  &  $\Sigma^2=\one$
\\
\hline
\hline
$0$ & $S^*\overline{P}S=P$ & $S^2=\one$ & $0$ & $0$ & $0$  & {\rm O}-$\ZM_2$
\\
\hline
$2$ & $S^*\overline{P}S=\one-P$ & $S^2=\one$ & {\rm O}-$\ZM_2$ & $0$ & $0$ & $0$
\\
\hline
$4$ & $S^*\overline{P}S=P$ & $S^2=-\one$  & $0$ & {\rm S}-$\ZM_2$ & $0$ & $0$
\\
\hline
$6$ & $S^*\overline{P}S=\one-P$ & $S^2=-\one$  & $0$ & $0$ & {\rm S}-$\ZM_2$ & $0$
\\
\hline
\end{tabular}
\end{center}

\end{proposi}

\noindent {\bf Proof.} First recall from Proposition~\ref{prop-IndTT'} that $\Ind(T)=0$. Let us first look at the entries in the column $d=1$. Set $F=\one-2E$ which is then unitary and satisfies both $\Sigma^*\overline{F} \Sigma=F$ and $\Sigma^*F^t \Sigma=F$ with $\Sigma^2=\one$. Hence the pairing $PFP-\one-P$ lies both in the column $d=2$ and $d=8$ of the table of Theorem~\ref{theo-indexlist}. Consequently, for $j=0$ the entry is both $0$ and $\ZM$, leading to the entry $0$. Similarly, one argues for $j=6$. For $j=2$, the entry is both $\ZM$ and O-$\ZM_2$, leading to O-$\ZM_2$ for $(j,d)=(2,1)$. For $j=4$ the entries are Q-$\ZM_2$ and Q-$2\,\ZM$, implying an entry $0$ because Q-$2\,\ZM$ actually means that the kernel is even dimensional. This concludes all cases of the column $d=1$. For column $d=5$ one can proceed in a similar manner using columns $d=4,6$ of Theorem~\ref{theo-indexlist}. Furthermore, these arguments applies also to the lines $j=0$ and $j=4$. Hence only remains to consider the four entries $(j,d)\in\{(2,3),\,(2,7),\,(6,3),\,(6,7)\}$. For $(j,d)=(2,3)$, let us note that $\Ker(T)=\Ker(P\imath(\one-2E)P+\one-P)$. Setting now $F=\imath(\one-2E)$ one has $\Sigma^*\overline{F}\Sigma=F$ and hence the entry $(j,d)=(2,4)$ allows to conclude. For $(j,d)=(6,7)$ one can proceed in the same manner. Next let us consider $(j,d)=(2,7)$. Using $[P,\Sigma]=0$ one finds $\Ker(T)=\Ker(P\Sigma(\one-2E)P+\one-P)$. Now $F=\Sigma(\one-2E)=-F^t$ and by Remark~\ref{rem-antisym} one concludes that $\Ker(T)$ is even dimensional so that the $\ZM_2$-index vanishes. Again the case $(j,d)=(6,3)$ works similarly.
\hfill $\Box$

\subsection{Pairing unitaries with unitaries}
\label{sec-FU}

\begin{proposi}
\label{prop-FU} 
Let $U$ and $F$ be two unitaries such that $T$ given in {\rm \eqref{eq-UniUni}} is a Fredholm operator. Then $\Ind_2(T)$ takes the following values:

\begin{center}
\begin{tabular}{|c|c c||c|c|c|c|}
\hline
& & & $d=2$ & $d=4$ & $d=6$ & $d=8$
\\
\hline
\multirow{2}{*}{$j$} & $[S,F]=0$ & $[\widehat{S},F]=0$ & $\Sigma^*F^t\Sigma=F$ & $\Sigma^* \overline{F}\Sigma=F$ & $\Sigma^*F^t\Sigma=F$ & $\Sigma^*\overline{F}\Sigma=F$
\\
 & $[\Sigma,P]=0$ &  & $\Sigma^2=\one$  & $\Sigma^2=-\one$ & $\Sigma^2=-\one$  &  $\Sigma^2=\one$
\\
\hline
$1$ & $S^*\overline{U}S=U$ & $S^2=\one$ & $0$ & $0$ & $0$  & {\rm O}-$\ZM_2$
\\
\hline
$3$ & $S^*U^tS=U$ & $S^2=-\one$ & {\rm Q}-$\ZM_2$ & $0$ & $0$ & $0$
\\
\hline
$5$ & $S^*\overline{U}S=U$ & $S^2=-\one$  & $0$ & {\rm S}-$\ZM_2$ & $0$ & $0$
\\
\hline
$7$ & $S^*U^tS=U$ & $S^2=\one$  & $0$ & $0$ &  {\rm R}-$\ZM_2$ & $0$
\\
\hline
\end{tabular}
\end{center}

\end{proposi}

\noindent {\bf Proof.} By Proposition~\ref{prop-IndT''T'''} one has $\Ind(T)=0$.  For $j=1,5$ the projection $P=\frac{1}{2}\binom{\one\;\;\;U }{U^*\;\;\one}$ satisfies
$$
\begin{pmatrix}
S & 0 \\ 0 & S
\end{pmatrix}^*
\overline{P}
\begin{pmatrix}
S & 0 \\ 0 & S
\end{pmatrix}
\;=\;
P
\;,
\qquad
\begin{pmatrix}
S & 0 \\ 0 & -S
\end{pmatrix}^*
\overline{P}
\begin{pmatrix}
S & 0 \\ 0 & -S
\end{pmatrix}
\;=\;
\one-P
\;.
$$
Consequently $P$ indeed satisfies the relations given in rows $j=1,5$ of Theorem~\ref{theo-indexlist} respectively and consequently one can copy these rows. For $j=3,7$ the relations are 
$$
\begin{pmatrix}
0 & S  \\  S & 0
\end{pmatrix}^*
\overline{P}
\begin{pmatrix}
0 & S \\ S & 0 
\end{pmatrix}
\;=\;
P
\;,
\qquad
\begin{pmatrix}
0 & S \\ -S & 0
\end{pmatrix}^*
\overline{P}
\begin{pmatrix}
0 & S \\ -S & 0
\end{pmatrix}
\;=\;
\one-P
\;.
$$
These are again the relations given in rows $j=3,7$ of Theorem~\ref{theo-indexlist}, with the sole difference that the two symmetry operators anti-commute. This can be fixed using the Cayley transformation $C=\frac{1}{\sqrt{2}}\binom{\one\;-\imath\one}{\one\;\;\;\imath\one}$ in the added fiber. The projections $P'=CPC^*=\frac{1}{2}\binom{\one-b\;\;-\imath a}{\imath a\;\;\;\;\one+b}$ with $a=\frac{1}{2}(U+U^*)$ and $b=\frac{1}{2\imath}(U-U^*)$ then satisfy
$$
\begin{pmatrix}
S & 0  \\  0 & S
\end{pmatrix}^*
\overline{P'}
\begin{pmatrix}
S & 0 \\ 0 & S
\end{pmatrix}
\;=\;
P'
\;,
\qquad
\begin{pmatrix}
0 & S \\ -S & 0
\end{pmatrix}^*
\overline{P'}
\begin{pmatrix}
0 & S \\ -S & 0
\end{pmatrix}
\;=\;
\one-P'
\;.
$$
Now the symmetries commute and are of the same type. Thus the index of $CT'C^*=P'\binom{F\; \;0}{0\;\;F}P'+\one-P'$ is given by the index in the rows $j=3,7$ of Theorem~\ref{theo-indexlist}. 
\hfill $\Box$

\vspace{.2cm}

Due to Proposition~\ref{prop-IndT''T'''} it is also possible to study instead of $T$ also the operator  $T''=UF+FU$. This lead to different index types. For example, for $(j,d)=(3,2)$ or $(j,d)=(7,6)$, one has $(S\Sigma)^*(T'')^t (S\Sigma)=T''$ with $(S\Sigma)^2=-\one$. Consequently, $T''$ is in the classifying space $\FM_2$ and has a non-vanishing index of type considered in Section~\ref{sec-indoddsym}.

\section{Index pairings for topological insulators}
\label{sec-overview}

\subsection{Topological insulators and their classification}
\label{sec-classification}

Let us consider a system of independent Fermions described by a bounded one-particle Hamiltonian $H=H^*$ acting on a complex Hilbert space $\Hh$ with complex conjugation $\Cc$. This Hamiltonian can have one or several of the following symmetries implemented by {\it commuting} symmetry operators $\RCH$, $\STR$ and $\SPH$ in the sense of Definition~\ref{def-SymmetryOp}:
\begin{align}
& \RCH^*H\RCH\;=\; - H\;,
& \RCH^2\,=\,\pm\one\;, & & \mbox{(chiral symmetry, CHS)}
\nonumber
\\
& \STR^*\overline{H}\STR\;=\; H\;,
& \STR^2\,=\,\pm\one \;, & & \mbox{(even/odd time-reversal symmetry, $\pm$TRS)}
\label{eq-symlist}
\\
& \SPH^*\overline{H}\SPH\;=\;- H\;,
& \SPH^2\,=\,\pm\one \;, & & \mbox{(even/odd particle-hole symmetry, $\pm$PHS)}
\nonumber
\end{align}
The CHS is sometimes also called a sublattice symmetry because that is the way it often appears in particular models, and a alternatively fermionic parity. The cases in which $H$ has no symmetry or only a CHS are called the complex classes (first two rows in Table~\ref{tab-class}; note that the sign of $\RCH^2$ is irrelevant there) and they will not be further considered here. The main focus is on the other $8$ so-called real classes because they invoke complex conjugation. There are $4$ cases with just one of the $4$ TRS and PHS, and there are $4$ combination of a TRS with a PHS. In these latter cases, their product $\RCH=\STR\SPH$ also induces a chiral symmetry. The symmetry classification into these $10$ classes has been known and widely used since the work of Altland and Zirnbauer \cite{AZ}. With each class is associated a Cartan-Altland-Zirnbauer (CAZ) label. This classification applies to metals, insulators and mesoscopic systems in the same manner.

\vspace{.2cm}

In the theory of topological insulators one now considers only Fermion systems for which the Fermi level lies either in a gap of $H$ or at least in a region of strong Anderson localization. The main new feature is that within several of the CAZ classes there are topologically different ground states which can be distinguished by so-called strong invariants. The possible values of these strong invariants are given in Table~\ref{tab-class}.  Integer quantum Hall systems ($d=2$ and $j=0$) are the best known examples with non-vanishing strong invariants, see \cite{BES} for a mathematical treatment. The first new topological insulators (with odd TRS and in dimension $d=2$) were theoretically discovered by Kane and Mele \cite{KM}, a complete list was found by Schnyder {\it et. al.} \cite{SRFL} and the periodic ordering of Table~\ref{tab-class} was put forward by Kitaev \cite{Kit}. In the same paper, Kitaev also showed that each entry in the lower part of the table has an interpretation in realm of Real $K$-theory \cite{Kar,Sc}. Let us briefly indicate how this goes. One considers $\RM^d=\RM^d_\tau$ as the momentum space on which complex conjugation is implemented by the involution $\tau(k)=-k$ for $k\in\RM^d$. This makes $C_0(\RM^d_\tau)$ into a Real C$^*$-algebra of which the $KR$-groups $KR_{j}(C_0(\RM^d_\tau))$, $j=0,\ldots,7$, can be defined. Each element of these $KR$-groups is interpreted as the class specified by a Fermi projection (of a translation invariant system) with a particular combination of the symmetries in \eqref{eq-symlist}. More precisely, the $(j,d)$th entry of Table~\ref{tab-class} is an invariant given by
\begin{equation}
\label{eq-invarform}
\mbox{\rm Inv}(j,d)
\;\in\;
KR_{j}(C_0(\RM^d_\tau))
\;=\;
\pi_{j-1-d}(O)
\;,
\end{equation}
where  the fundamental groups of the stabilized orthogonal group $O$ are $8$-periodic and given by \eqref{tab-HomO}. The second equality in \eqref{eq-invarform} can be found in textbooks \cite{Kar,Sc} and has been explained in various more recent works in the context of topological insulators \cite{SCR,FM,Tan,KZ}. The formula \eqref{eq-invarform} indeed explains the $8$ periodicity of the lower part of  Table~\ref{tab-class} both in $d$ and $j$. For periodic rather than translation invariant systems, the invariants take values in the larger group $KR_{j}(C_0(\TM^d_\tau))$. The supplementary elements are then the so-called weak invariants \cite{Kit} which allow to further distinguish ground states with equal strong invariants.

\vspace{.2cm}

From now on, it will be supposed that the Hamiltonian $H$ acts on the tight-binding Hilbert space $\ell^2(\ZM^d)\otimes\CM^N$ with $N$-dimensional fibers that the symmetry operators $\RCH$, $\STR$ and $\SPH$ act on these fibers only. Furthermore a Fermi level $\mu\in\RM$ is given. In all cases with either a CHS or a PHS it is supposed to be $\mu=0$.  It specifies the Fermi projection $P=\chi(H\leq \mu)$ as the spectral projection of $H$ on energies below the Fermi level.

\begin{table}
\begin{center}

\begin{tabular}{|c|c|c|c||c||c|c|c|c|c|c|c|c|}
\hline
$j$ & $\!\!$TRS$\!\!$ & $\!\!$PHS$\!\!$ & $\!\!$CHS$\!\!$ & CAZ & $\!d=0,8\!$ & $\!d=1\!$ & $\!d=2\!$ & $d\!=3\!$ & $\!d=4\!$ & $\!d=5\!$ & $\!d=6\!$ & $\!d=7\!$
\\\hline\hline
$0$ & $0$ &$0$&$0$& A  & $\ZM$ &  & $\ZM$ &  & $\ZM$ &  & $\ZM$ &  
\\
$1$& $0$&$0$&$ 1$ & AIII & & $\ZM$ &  & $\ZM$  &  & $\ZM$ &  & $\ZM$
\\
\hline\hline
$0$ & $+1$&$0$&$0$ & AI &  $\ZM$ & &  & & $2 \, \ZM$ & & $\ZM_2$ & ${\ZM_2}$
\\
$1$ & $+1$&$+1$&$1$  & BDI & $\ZM_2$ &$\ZM$  & &  &  & $2 \, \ZM$ & & $\ZM_2$
\\
$2$ & $0$ &$+1$&$0$ & D & $\ZM_2$ & ${\ZM_2}$ & $\ZM$ &  & & & $2\,\ZM$ &
\\
$3$ & $-1$&$+1$&$1$  & DIII &  & $\ZM_2$  &  $\ZM_2$ &  $\ZM$ &  & & & $2\,\ZM$
\\
$4$ & $-1$&$0$&$0$ & AII & $2 \, \ZM$  & &  $\ZM_2$ & ${\ZM_2}$ & $\ZM$ & & &
\\
$5$ & $-1$&$-1$&$1$  & CII & & $2 \, \ZM$ &  & $\ZM_2$  & $\ZM_2$ & $\ZM$ & &
\\
$6$ & $0$ &$-1$&$0$ & C&  &  & $2\,\ZM$ &  & $\ZM_2$ & ${\ZM_2}$ & $\ZM$ &
\\
$7$ & $+1$&$-1$&$1$  &  CI &  & &   & $2 \, \ZM$ &  & $\ZM_2$ & $\ZM_2$ & $\ZM$
\\
[0.1cm]
\hline
\end{tabular}
\end{center}
\caption{{\it List of symmetry classes ordered by {\rm TRS}, {\rm PHS} and {\rm CHS} as well as the {\rm CAZ} label. Then follow the strong invariants in dimension $d=0,\ldots,8$. }}
\label{tab-class}
\end{table}

\subsection{Invariants for zero-dimesional systems}
\label{sec-Invd=0}

The main aim of the present work is to calculate the invariant $\Inv(j,d)$ not as a $K$-group element as in \eqref{eq-invarform}, but rather as numerical invariant from an index pairing for systems submitted to the basic symmetries.  As a warm-up let us consider the case of dimension $d=0$, following \cite{Kit,HL}, see also \cite{Lor}. Here the Hamiltonian $H$ is merely a finite dimensional self-adjoint matrix (the size of this matrix is the dimension of fiber over a point constituting the $0$-dimensional system) which satisfies some of the fundamental symmetries \eqref{eq-symlist} implemented by matrices $\STR$ and $\SPH$. The Fermi level is supposed to lie in a gap of the spectrum of $H$. For CAZ classes A, AI and AII the invariant distinguishing different systems is simply the signature of $H-\mu\one$ given by $\Inv(j,0)=\Tr(P)-\Tr(\one-P)$ for $j=0,4$. This is clearly a homotopy invariant which can only change when $\mu$ is an eigenvalue of $H$ (so that the system is not an insulator any more). Moreover, in the case of odd TRS (Class AII) a Kramers degeneracy argument indeed shows that the signature is always even, leading to the $2\,\ZM$ entry for $(j,d)=(5,0)$. If one considers a Hamiltonian with PHS, then the Fermi projection $P=\chi(H\leq 0)$ satisfies
$$
\SPH^*\overline{P}\SPH
\;=\;
\one-P
\;,
$$
namely $P$ is Lagrangian w.r.t. the symmetry $\SPH$. Taking the trace of the identity immediately implies that the signature vanishes. A similar argument also applies for chiral systems, in particular Class AIII. Hence only remains to explain the two (secondary) $\ZM_2$-invariants appearing in Classes D and BDI. In typical models (stemming from second quantized operators which are quadratic in the creation and annihilation operators), the even PHS and Hamiltonian are of the form
$$
H
\;=\;
\begin{pmatrix}
h & \Delta \\ \Delta^* & -\overline{h}
\end{pmatrix}
\;,
\qquad
\SPH\;=\;
\begin{pmatrix}
0 & \one \\ \one & 0
\end{pmatrix}
\;.
$$
Hence $\SPH=\sigma_1\otimes\one$. Now the PHS $\SPH \overline{H}\SPH =-H$ is equivalent to the BdG equation $\Delta^t=-\Delta$. The invariant is best defined in the Majorana representation obtained after Cayley transformation in the grading of $\SPH$:
\begin{equation}
H_\Maj
\;=\;
C^t\,H\,\overline{C}
\;,
\qquad
C\;=\;
\frac{1}{\sqrt{2}}
\begin{pmatrix}
\one & -\,\imath\,\one \\
\one & \imath\,\one
\end{pmatrix}
\;.
\label{eq-BdGMaj}
\end{equation}
Then the PHS becomes $(C^*\SPH  \overline{C})^*\overline{H_\Maj}  (C^*\SPH \overline{C})=-  H_\Maj$ so that the new symmetry operator is $C^*\SPH \overline{C}=C^*\sigma_1  \overline{C}=\one$. Hence 
$$
H_\Maj
\;=\;
H_\Maj^*
\;=\;
-\,\overline{H_\Maj}
\;=\;
-\,H_\Maj^t
\;,
$$
Thus $H_\Maj$ is a purely imaginary and antisymmetric matrix. Its matrix entries can readily expressed in terms of the real and imaginary parts of the matrix entries of $H$.  Now one has, as for every real antisymmetric matrix, $\det(\imath H_\Maj)=\mbox{\rm Pf}(\imath H_\Maj)^2$ and  the $\ZM_2$-invariant is defined as the sign of the Pfaffian $\Inv(2,0)=\mbox{\rm sgn}(\mbox{\rm Pf}(\imath H_\Maj))\in\ZM_2$. For systems of Class BDI ($j=1$) which, moreover, have an even TRS, this Pfaffian invariant $\Inv(1,0)=\mbox{\rm sgn}(\mbox{\rm Pf}(\imath H_\Maj))\in\ZM_2$ is still well-defined and may be non-trivial. This completes our discussion of $0$-dimensional systems.

\subsection{Reordering of the symmetries of the Hamiltonian}
\label{sec-Ham}

As discussed in Section~\ref{sec-overview}, the Hamiltonian is supposed to have one or two of the commuting physical symmetries $\STR$ and $\SPH$ given by \eqref{eq-symlist}. Two of these symmetries then induce a chiral symmetry $\RCH=\STR\SPH$. The first task is to order these symmetries in the way they appear in the periodic table (Table~\ref{tab-class}). This is written out in the following table: 

\vspace{.1cm} 

\begin{equation}
\label{tab-HamSym}
\begin{tabular}{|c||c|c|c|c|c|c|c|c|}
\hline
$j$ & $0=8$ & $7$ & $6$ & $5$ & $4$ & $3$ & $2$ & $1$
\\
\hline
CAZ
& AI & CI & C & CII & AII & DIII & D & BDI
\\
\hline
\hline
$S^*\overline{P}S=$ & $P$  & $\one-P$   & $\one-P$ & $P$   & $P$ & $\one-P$   & $\one-P$ & $P$    
\\
$S$ & $+$TRS  &  $-$PHS   & $-$PHS & $-$TRS   & $-$TRS &  $+$PHS  & $+$PHS & $+$TRS    
\\
\hline
$\widehat{S}^*\overline{P}\widehat{S}=$ &   & $P$   &  & $\one-P$   & & $P$   &  & $\one-P$    
\\
$\widehat{S}$ &   & $+$TRS  &  &  $-$PHS  &  & $-$TRS  &  & $+$PHS    
\\
\hline
$R^2=$ & & $-\one$ & & $\one$  & & $-\one$ & & $\one$
\\
$R^*{P}R=$ &  & $\one-P$ &  & $\one-P$ & & $\one-P$ &  & $\one-P$  
\\
\hline
\end{tabular}
\end{equation}

\vspace{.1cm} 

\noindent Here $S$ and $\widehat{S}$ are either $\STR$ or $\SPH$ and $R=S\widehat{S}=\RCH$. Let us note that it is not possible to choose $S=\STR$ or $S=\SPH$ throughout for all $j$ because both a single TRS and a single PHS appear as single symmetries (for $j$ even). On the other hand for odd $j$ we made a particular (arbitrary) choice as to what is $S$ and what is $\widehat{S}$. It turns out that this ordering is obtained precisely by reversing the ordering of the Dirac operator in Section~\ref{sec-Dirac} below (which is inherited from the Clifford group structure), namely $j=9-d$ for $d=1,\ldots,8$.

\vspace{.2cm}

For odd $j$, one needs to reduce out the Fermi projection to a unitary \cite{RSFL,PS} by going to a basis in which the chiral symmetry is proportional to the third Pauli matrix $\sigma_3$. By Proposition~\ref{prop:unitaries-commute-1} and the hypothesis that the eigenspaces of all symmetry operators are of equal dimension, there exists a real unitary basis transformation $O$ such that

\vspace{.1cm} 

\begin{center}

\begin{tabular}{|c||c|c|c|c|}
\hline
$j$ & $7$  & $5$  & $3$  & $1$ 
\\
\hline
$O^\ast R O=$ & $\imath \sigma_2\otimes\one$ & $\sigma_3\otimes\one$  & $ \imath \sigma_2\otimes\one$ & $\sigma_3\otimes\one$  
\\
 $O^\ast S O=$ & $\imath \sigma_2 \otimes\one$ & $\one\otimes\imath\sigma_2$  & $ \imath\sigma_2 \otimes \imath\sigma_2$ & $\one\otimes\one$ 
\\
\hline
$\imath\;C O^\ast R O C^\ast=$ & $\sigma_3\otimes\one$  & & $\sigma_3\otimes\one$ & 
\\
$\imath\;C O^\ast S O C^t=$ & $ \imath\sigma_2 \otimes \one$  & & $\imath\sigma_2 \otimes \imath\sigma_2$   &
\\
\hline
\end{tabular}
\end{center}

\vspace{.1cm} 

\noindent In the last two columns a supplementary Cayley transformation was carried out so that the chiral symmetry is proportional to the Pauli matrix $\sigma_3$. Let us point out that the Cayley transform is taken in the first factor, and defined exactly as in \eqref{eq-BdGMaj} and Section~\ref{sec-CGLie} so that all identities stated there can be used. The effect of the Cayley transform is to change the symmetry operators from commuting to anti-commuting, just as in the proof of Proposition~\ref{prop-FU} and Appendix~\ref{sec-CGLieG}. This results from the fact that the TRS and PHS have to be transformed as $S\mapsto CSC^t$, while the chiral symmetry as $R\mapsto CRC^*$, see again Appendix~\ref{sec-CGLieG}. In the new basis, one can now reduce out $P$ to a unitary $U$. The symmetry $S$, which is a TRS for $j=1,5$ and a PHS for $j=3,7$, now implies that the unitary $U$ has a symmetry as listed in the following table:

\vspace{.1cm} 

\begin{equation}
\label{tab-PSymOdd}
\begin{tabular}{|c||c|c||c||c|c|}
\hline
$j$ &  $5$  & $1$ &  & $7$ & $3$
\\
\hline
$O^\ast(2P-\one)O=$       & $\binom{0\;\;U}{U^*\,0}$   &   $\binom{0\;\;U}{U^*\,0}$
 & $ C O^\ast(2P-\one)O C^\ast =$ & $\binom{0\;\;U}{U^*\,0}$ & $\binom{0\;\;U}{U^*\,0}$
\\
[0.06cm]
$U=$    & $\sigma_2^*\overline{U}\sigma_2$ &     $\overline{U}$  
 &$U=$ &  $U^t$ &  $\sigma_{2}^* U^t \sigma_{2}$
\\
\hline
\end{tabular}
\end{equation}

\vspace{.1cm} 

\noindent In conclusion, for every chiral system  (namely $j$ odd) a unitary operator $U$ is needed to specify the Fermi projection. This unitary unitary inherits symmetries from $P$. Before going on, let us point out that  arbitrary unitary transformations of the Hamiltonian may change the unitary $U$, {\it e.g.} clearly for $j=1,5$
$$
\begin{pmatrix} 
\one & 0 \\ 0 & U 
\end{pmatrix}
O^*(2P-\one)O
\begin{pmatrix} 
\one & 0 \\ 0 & U 
\end{pmatrix}^*
\;=\;
\begin{pmatrix} 
0 & \one  \\ \one & 0 
\end{pmatrix}
\;,
$$
which appears to radically change the $K$-theoretic data. Indeed, the unitary transformation shifts the two chiral components w.r.t. each other by $U$ and this does change the model in an essential manner and is therefore not allowed. The basis transformations $O$ and $C$ in \eqref{tab-PSymOdd} are going to be local for thight-binding models on $\ell^2(\ZM^d)\otimes\CM^N$ considered below, namely they commute with the position operators.

\subsection{The Dirac operator and its symmetries}
\label{sec-Dirac}

In the last section, the Hamiltonian together with its Fermi level lead to a projection $P$ on $\ell^2(\ZM^d) \otimes \CM^N$ and for odd $j$ a unitary $U$ on $\ell^2(\ZM^d) \otimes \CM^{\frac{N}{2}}$. This data will constitute the $K$-theoretic input to the index pairings in Section~\ref{sec-FredTopIns} below. The $K$-homological part of the pairing is deduced from the (unbounded self-adjoint) Dirac operator
\begin{equation} 
\label{eq-diracop}
D
\; =\; 
\sum_{j=1}^d X_j \otimes \one \otimes \Gamma_j  
\,,
\end{equation}
which acts on the Hilbert space $\Hh=\ell^2(\ZM^d) \otimes \CM^N \otimes \CM^{d'} $ where $d' = 2^{\lfloor \frac{d}{2} \rfloor}$. Here $X_1,\ldots,X_d$ are the components of the position operator on $\ell^2(\ZM^d)$ and the $\Gamma_1,\ldots,\Gamma_d\in\CM^{d'\times d'}$ are anti-commuting, square to $\one$ and are such that $\Gamma_{2n}$ is imaginary and $\Gamma_{2n+1}$ is real. The latter fact allows to implement complex conjugation in a convenient manner. In Appendix~\ref{app-Cd}, it is shown how these $\Gamma_n$'s are constructed as a particular irreducible representation of the Clifford algebra $C_d$. The operator $D$ defined in \eqref{eq-diracop} may not look like the usual Dirac operator, but after a discrete Fourier transform $\Ff$ it takes the more familiar form 
$$
\Ff\,D\,\Ff^*
\;=\;
\sum_{j=1}^d \imath\partial_{k_j} \otimes \one \otimes \Gamma_j
\;,
$$ 
acting on the Hilbert space $L^2(\TM^d)\otimes\CM^{N}\otimes\CM^{d'}$ over the Brillouin torus. Due to Proposition~\ref{prop-SigmaUnique}, the Dirac operator hence inherits from the $\Gamma$'s the following symmetries:
\begin{equation}
\label{eq-DiracSymShort}
\Sigma^*\,\overline{D}\,\Sigma
\;=\;
\kappa\,D
\;,
\qquad
\Omega^*\,D\,\Omega\;=\;
-\,D
\;,
\qquad
\widehat{\Sigma}^*\,\overline{D}\,\widehat{\Sigma}\;=\;
-\kappa\,D
\;,
\end{equation}
where the latter two are only given for even $d$. Here $\Sigma$, $\Omega$ and $\widehat{\Sigma}$ are symmetry operators which are defined in Appendix~\ref{app-Cd} and $\kappa=(-1)^{\lfloor \frac{d}{2}\rfloor}$. In particular, $\Omega\Sigma=\Sigma\Omega$ for $d=4,8$ and $\Omega\Sigma=-\Sigma\Omega$ for $d=2,6$. Using the terminology introduced for the Hamiltonian in Section~\ref{sec-overview}, $\kappa=1$ leads to a PHS for $\Sigma$ and a TRS for $\widehat{\Sigma}$, while $\kappa=-1$ implies that $\Sigma$ is a TRS and $\widehat{\Sigma}$ a PHS. Pending on the signs of $\Sigma^2$ and $\widehat{\Sigma}^2$ these symmetries are even or odd. All these properties together allow to interpret $D$ as a $KR$-cycle, see Section~\ref{sec-NCG}. Associated to $D$ is again a spectral projection
$$
E
\;=\;
\chi(D> 0)\;+\;e_0\otimes\one\otimes\chi(\Gamma_1>0)
\;,
$$
where $e_0$ is the projection on the state $\ell^2(\ZM^d)$ over the origin in $\ZM^d$. The projection $E$ will be called Hardy projection as in dimension $d=1$ it projects on the Hardy space of positive frequencies. Now the properties \eqref{eq-DiracSymShort} and Proposition~\ref{prop-SigmaUnique} can be summarized in the following table:

\vspace{.1cm}

\begin{equation}
\label{tab-DiracSym}
\begin{tabular}{|c||c|c|c|c|c|c|c|c|}
\hline
$d\,$mod$\,8$ & $1$ & $2$ & $3$ & $4$ & $5$ & $6$ & $7$ & $8$
\\
\hline\hline
$\Sigma^*\,\overline{E}\,\Sigma=$ & $E$  & $\one-E$   & $\one-E$ & $E$   & $E$ & $\one-E$   & $\one-E$ & $E$    
\\
$\Sigma^2=$ & $\one$ & $-\one$ & $-\one$ &  $-\one$  &  $-\one$  & $\one$ & $\one$ &  $\one$ 
\\
\hline
$\widehat{\Sigma}^*\,\overline{E}\,\widehat{\Sigma}=$ &  & $E$ &  & $\one-E$ & & $E$ &  & $\one-E$  
\\
$\widehat{\Sigma}^2=$ & & $\one$ & & $-\one$ & & $-\one$ & & $\one$
\\
[0.1cm]
\hline
$\Omega^*E\Omega=$ &  & $\one-E$ &  & $\one-E$ & & $\one-E$ &  & $\one-E$  
\\
[0.02cm]
\hline
$\!2\,O^*EO-\one=\!$ & &  $\binom{0\;\;F}{F^*\,0}$ & &  $\binom{0\;\;F}{F^*\,0}$ & & $\binom{0\;\;F}{F^*\,0}$ &  & $\binom{0\;\;F}{F^*\,0}$  
\\
[0.07cm]
$F=$ & &  $F^t$ & & $\sigma_2^*\overline{F}\sigma_2$  & & $-F^t$ &  & $\sigma_1^*\overline{F}\sigma_1 $  
\\
\hline
\end{tabular}
\end{equation}

\vspace{.1cm}

\noindent In the last two rows the orthogonal basis change $O$ of Proposition~\ref{prop-GammaSigmaNormal} is used, so that the chiral symmetry allows to deduce a unitary $F$ from $E$ which then has the symmetries indicated in an analogous manner as the Fermi projection $P$ leads to a unitary $U$ in Section~\ref{sec-Ham}. In the index theorems for the even dimensions, it will tacitly be assumed below that this representation is chosen. These considerations already conclude the analysis of the symmetries of the Dirac operator, the Hardy projection and the Dirac phase, namely the $K$-homological part of the pairing.

\subsection{Fredholm operators for topological insulators}
\label{sec-FredTopIns}

In this section the numerical invariants $\Inv(j,d)$ for higher dimensions $d\geq 1$ are going to be calculated as index pairings of $P$ and $U$ paired with $E$ and $F$, all given by the tables \eqref{tab-HamSym}, \eqref{tab-PSymOdd} and \eqref{tab-DiracSym}. These index pairings are listed in \eqref{tab-IndPair0}, namely one sets
\begin{equation}
\label{eq-pairingdef}
T
\;=\;
\left\{
\begin{array}{cc}
PFP\,+\one-P\;, &d\; \mbox{\rm even}\;,
\\
E\,UE\,+\,\one-E\;, &  \;\;\;\;\;\;\;\;\;\; \;\;\;\;\; d\; \mbox{\rm odd and }j\;\mbox{\rm odd}\;,
\\
E\,(2P-\one)\,E\,+\,\one-E\;, &  \;\;\;\;\;\;\;\;\;\; \;\;\;\;\;\;\; d\; \mbox{\rm odd and }j\;\mbox{\rm even}\;,
\end{array}
\right.
\end{equation}
The first two operators act on $\ell^2(\ZM^d)\otimes\CM^{\frac{Nd'}{2}}$, the last one on $\ell^2(\ZM^d)\otimes\CM^{{Nd'}}$. Let us note the symmetry of $T$  upon exchange $(P,F)\leftrightarrow (E,U)$ under which the $K$-theoretic and $K$-homological parts exchange roles. Due to Proposition~\ref{prop-IndT''T'''} this symmetry also holds for the case of odd $d$ and even $j$. Of course, the following will be crucial:

\vspace{.2cm}

\noindent {\bf Standing Hypothesis} {\it $T$ is a Fredholm operator on $\Hh=\ell^2(\ZM^d)\otimes\CM^{\frac{Nd'}{2}}$.}

\vspace{.2cm}

\noindent It is shown in \cite{BES,PLB,PS} that this Fredholm property actually holds when $\mu$ lies in a gap or, more generally, in a region of dynamical Anderson localization for covariant random models. Now \eqref{eq-pairingdef} are the pairings considered in Section~\ref{sec-indsym}. The conclusions are resumed in the following theorem.

\begin{theo}
\label{theo-topinslist}
Each of the entries of the periodic {\rm Table~\ref{tab-class}} of topological insulators can be understood as $\ZM$, $2\,\ZM$ or $\ZM_2$-index theorem in the sense of {\rm Theorem~\ref{theo-indexlist}} associated to the Fredholm operators {\rm \eqref{eq-pairingdef}}. In the terminology of {\rm Section~\ref{sec-indsym}}, the type {\rm R,\,Q,\,S} or {\rm O} of the index theorem is given in the following table:
\begin{center}

\begin{tabular}{|c||c|c|c|c|c|c|c|c|}
\hline
   & $d=1$ & $d=2$ & $d=3$ & $d=4$ & $d=5$ & $d=6$ & $d=7$ & $d=8$
\\\hline\hline
$j=0$    & &  &  & {\rm R}-$2 \, \ZM$ & & {\rm R}-$\ZM_2$ &  {\rm O}-$\ZM_2$ &$\ZM$
\\
$j=1$   & $\ZM$ &  & & & {\rm Q}-$2\,\ZM$ & & {\rm O}-$\ZM_2$ & {\rm O}-$\ZM_2$
\\
$j=2$   &  {\rm R}-$\ZM_2$ &  $\ZM$ &  & & &  {\rm O}-$2\,\ZM$ &   & {\rm O}-$\ZM_2$
\\
$j=3$   &  {\rm R}-$\ZM_2$ &  {\rm Q}-$\ZM_2$ & $\ZM$ & & & & {\rm O}-$2 \, \ZM$ &
\\
$j=4$  &  &  {\rm Q}-$\ZM_2$  &  {\rm S}-$\ZM_2$ & $\ZM$ & & & & {\rm Q}-$2 \, \ZM$
\\
$j=5$   & {\rm R}-$2\,\ZM$ &  &  {\rm S}-$\ZM_2$ & {\rm S}-$\ZM_2$ & $\ZM$ & & &
\\
$j=6$   &   &  {\rm S}-$2 \, \ZM$ &  & {\rm S}-$\ZM_2$ & {\rm Q}-$\ZM_2$ & $\ZM$& &
\\
$j=7$   &  & & {\rm S}-$2 \, \ZM$ & & {\rm Q}-$\ZM_2$ & {\rm R}-$\ZM_2$ &  $\ZM$ &
\\
[0.1cm]
\hline
\end{tabular}
\end{center}

\end{theo}

\noindent {\bf Proof.} For a given $(j,d)$ one reads of the symmetries of $P$ or $U$ and $E$ or $F$ from the tables \eqref{tab-HamSym}, \eqref{tab-PSymOdd} amd \eqref{tab-DiracSym}. A careful check then allows to reduce all entries of the table above to those of Theorem~\ref{theo-indexlist} or Proposition~\ref{prop-PE}. In particular, the sub-table obtained by considering only even $d$ is precisely the table of Theorem~\ref{theo-indexlist}. 
\hfill $\Box$

\vspace{.2cm}

Let us note that there is little difference between $\ZM$- and $2\,\ZM$-indices if one is merely interested in distinguishing ground states. There is, however, a meaning associated to the evenness of indices. For example \cite{DS2}, in dimension $d=2$ a BdG Hamiltonian (Class D or C, $j=2$ or $j=6$) has a Majorana zero mode attached to a vortex defect. While indeed there is such a zero mode in Class D, there is none for a Class C system because the index is always even. This is also reflected by the fact that going $2$ dimensions to the left in the periodic table (the vortex point defect is an effectively zero-dimensional system) one finds a $\ZM_2$-index for $(j,d)=(2,0\simeq 8)$, but no entry for $(j,d)=(6,0\simeq 8)$.

\subsection{Examples}
\label{sec-example}

The aim is here not to produce an exhaustive list of examples, but rather to present some general recipes  for the construction of non-trivial models with TRS or PHS from models in the unitary classes (Class A and AIII). In particular, the focus is on  topological insulators with $\ZM_2$-indices in low dimensions which are of experimental relevance \cite{HK}, namely Class D in dimension $d=1$ (Kitaev chain \cite{Kit0}), Class AII in dimension $d=2$ (Kane-Mele model \cite{KM}) and Class AII in $d=3$ (Fu-Kane-Mele strong topological insulator \cite{FKM}).

\subsubsection{Examples in dimension $d=1$}

The infinite and clean Kitaev chain is described by a Hamiltonian on $\ell^2(\ZM)\otimes\CM^2$ given by
$$
H
\;=\;
\frac{1}{2}\,
\begin{pmatrix}
V+V^*+2\mu & \imath(V-V^*) \\
\imath(V-V^*) & -(V+V^*+2\mu)
\end{pmatrix}
\;,
\qquad
\SPH\;=\;
\begin{pmatrix}
0 & \one\\
\one & 0
\end{pmatrix}
\;.
$$
Here $V$ denotes the left shift and $\mu\in\RM$ is a chemical potential. This model has an even PHS with $\SPH=\sigma_1$ and an even TRS $\STR=\sigma_3$ where $\sigma_1$ and $\sigma_3$ are the Pauli matrices acting on the fiber. Consequently, the model lies in the Class BDI and, according to Theorem~\ref{theo-topinslist}, has a well-defined $\ZM$-index (this is independent of the fact that the two symmetries actually anti-commute, upon Cayley transform the become commuting without changing the class). For the calculation of the $\ZM$-index, one has to diagonalize the chiral symmetry $\RCH=\SPH\STR=\imath\sigma_2$ which is realized by the Cayley transform $C$ defined in \eqref{eq-BdGMaj}, that is $C\RCH C^*=-\imath\sigma_3$. Then
$$
C\, H\, C^*
\;=\;
\begin{pmatrix}
0 & V+\mu \\
V^*+\mu & 0
\end{pmatrix}
\;,
\;\;\;
C\, (2P-\one)\, C^*
\;=\;
\begin{pmatrix}
0 & \!\!\!(V+\mu)|V+\mu|^{-1} \\
(V^*+\mu)|V^*+\mu|^{-1} & 0
\end{pmatrix}
.
$$
Now the unitary $U=(V+\mu)|V+\mu|^{-1}$ and Hardy projection $E=\chi(X\geq 0)$ allow to define a Fredholm operator $T=EUE+\one-E$ for $\mu\not\in\{-1,1\}$. For  $\mu\in\{-1,1\}$ the gap is closed and $T$ is not Fredholm.  Now for $|\mu|<1$ one has $\Ind(T)=\Ind(EUE|_{\Ran(E)})=\Ind(E(V+\mu)E|_{\Ran(E)})=1$ because $EVE|_{\Ran(E)}$ is the unilateral shift on $\ell^2(\NM)$, while for $|\mu|>1$ one has $\Ind(T)=0$. 


\vspace{.2cm}

Now let us add a supplementary term to the Hamiltonian which breaks the TRS, but not the even PHS. Then $H$ has no chiral symmetry any more and lies in Class D. Hence by Theorem~\ref{theo-topinslist} the $\ZM_2$-index $\Ind_2(E(2P-\one)E+\one-E)$ is still well-defined, and it can be read off the formula above that it is actually equal to the non-trivial value $1$ for $|\mu|<1$. Let us point out that this $\ZM_2$-index is stable under (random) perturbations of the Hamiltonian which conserve the even PHS. This $\ZM_2$-index is equal to $\Ind_2(EHE+\one-E)$, namely the multiplicity of the zero modes of the half-space Hamiltonian $EHE$. This provides an interesting physical interpretation of the $\ZM_2$-index.

\subsubsection{Examples in even dimension}
\label{sec-evenexample}

In even dimensions $d$, there are models of Class A (no symmetry at all) which have non-vanishing (higher even) Chern numbers \cite{RSFL} which by an index theorem \cite{PLB} are equal to the index of a Fredholm operator described next. Let the Hamiltonian $h=h^*$ act on $\ell^2(\ZM^d)\otimes \CM^N$ and let $p=\chi(h\leq \mu)$ be the Fermi projection associated to a Fermi energy lying in a region of dynamical localization. If $F$ is the Dirac phase as constructed in Section~\ref{sec-Dirac}, then the Chern number is equal to $\Ind(pFp+\one-p)$. From such a model let us now construct further non-trivial models with symmetries, similar as in \cite{DS2} and inspired by the construction of the Kane-Mele model from two copies of the Haldane model. The new Hamiltonian on $\Hh=\ell^2(\ZM^2)\otimes \CM^N\otimes \CM^2$ is of the form
\begin{equation}
\label{eq-KMdouble}
H
\;=\;
\begin{pmatrix} h & g \\ g^* & \overline{h}
\end{pmatrix}
\;,
\end{equation}
in the grading of the spin degree of freedom $\CM^2$. It then has an odd TRS with $\STR=\imath\sigma_2$ whenever $g^t=-g$. The next aim is to calculate the invariant for $H$ from that of $h$. For vanishing  $g$, the Fermi projection $P=\chi(H\leq \mu)$ is given by $P=\binom{p\;\;0}{0\;\;\overline{p}}$. Therefore
$$
PFP
\;=\;
\begin{pmatrix}
p\,F\,p & 0 \\
0 & \overline{p}\,F\,\overline{p}
\end{pmatrix}
\;.
$$
Now suppose that $\Ind(pFp+\one-P)$ is odd. Therefore, if $\dim(\Ker(pFp))$ is odd (resp. even), then $\dim(\Ker(pF^*p))=\dim(\Ker(p\,\overline{F}\,p))=\dim(\Ker(\overline{p}\,F\,\overline{p}))$ is even (resp. odd). It follows that $\dim(\Ker(PFP))$ is indeed odd so that $\Ind_2(PFP+\one-P)=1$.  If now $g=-g^t$ is added homotopically and $d=2$, then this non-trivial $\ZM_2$-index is conserved by homotopy invariance. If $g$ is chosen to be the Rashba coupling and $h$ the Haldane Hamiltonian, then the model obtained in this manner is precisely the Kane-Mele model. In a similar way, one can produce a  model in $d=6$ with even TRS and non-trivial $\ZM_2$-invariant.

\vspace{.2cm}

Now let us indicate the changes needed to construct a topologically non-trivial model with a PHS. One rather begins from
\begin{equation}
\label{eq-PHSdouble}
H'
\;=\;
\begin{pmatrix} h & f \\ f^* & -\overline{h}
\end{pmatrix}
\;,
\end{equation}
and then imposes $f=-f^t$ for an even PHS $\SPH=\sigma_1$, or $f=f^t$ for an odd PHS $\SPH=\imath\sigma_2$. The system with even PHS is of particular interest in dimension $d=8$ where a $\ZM_2$-topological insulator can be constructed, and the odd PHS is of relevance in $d=4$. The construction of systems with two simultaneous symmetries is an extension left to the reader.

\subsubsection{Examples in odd dimension}
\label{sec-oddexample}

The procedure of Section~\ref{sec-evenexample} can be modified in order to produce topological models in odd dimension $d$ which have two commuting symmetries, namely $j=1,3,5,7$. For that purpose, one starts from a Hamiltonian $h=h^*$ on $\ell^2(\ZM^d)\otimes\CM^N\otimes \CM^2$ with CHS $\sigma_3 h\sigma_3=-h$. In the grading of $\sigma_3$ it is of the form $h=\binom{0\;\;\;a}{a^*\;0}$, and under a localization hypothesis it has an index $\Ind(EaE+\one-E)$ which is connected to odd higher Chern numbers \cite{RSFL} by an index theorem \cite{PS}. Now one can construct systems with TRS similar as in \eqref{eq-KMdouble} by
%
$$
H
\;=\;
\begin{pmatrix} h & g \\ g^* & \sigma_1\overline{h}\sigma_1
\end{pmatrix}
\;,
$$
%
This Hamiltonian has a CHS $\RCH=\sigma_3\otimes\one$ as long as $\sigma_3 g\sigma_3=-g$. Furthermore, it has an odd TRS $\STR=\sigma_1\otimes\imath\sigma_2$ whenever $g=-\sigma_1g^t\sigma_1$, that is $g=\binom{0\;\;b}{c\;\;0}$ with anti-symmetric $b$ and $c$. Hence it is in Class CII, $j=5$. Now one checks
$$
W^*\,H\,W
\;=\;
\begin{pmatrix}
0 & 0 & a & b \\
0 & 0 & c^* & a^t \\
a^* & c & 0 & 0 \\
b^* & \overline{a} & 0 & 0
\end{pmatrix}
\;,
\qquad
W
\;=\;
\begin{pmatrix}
1 & 0 & 0 & 0 \\
0 & 0 & 1 & 0 \\
0 & 1 & 0 & 0 \\
0 & 0 & 0 & 1
\end{pmatrix}
\;.
$$
The unitary $U$ of \eqref{tab-PSymOdd} is the phase of the invertible $\binom{a \;\; b}{  c^* \; a^t}$, which is the upper right entry of $W^*HW$. For $b$ and $c$ small, one thus has $\Ind(T)=0$ for $T=EUE+\one-E$, however, it is possible that there is a secondary $\ZM_2$-invariant. For $d=3$, this $\ZM_2$-invariant is non-trivial provided that $\Ind(EaE+\one-E)$ is odd by an argument similar to the one in Section~\ref{sec-evenexample}.

\vspace{.2cm}

Finally let us construct a three-dimensional model in Class AII which has a non-trivial $\ZM_2$-invariant. This will be achieved by the Hamiltonian given in equation (82) of \cite{RSFL}.  Let $(\Gamma_j)_{j=1,\ldots,5}$ be an irreducible representation of the complex Clifford algebra $C_5$ on $\CM^4$, exactly as given in Section~\ref{sec-CGrep}. Consider the following Hamiltonian on the Hilbert space $\ell^2(\ZM^3)\otimes \CM^4$:
$$
H
\;=\;
\sum_{j=1}^3\frac{1}{2\imath}(V_j-V_j^*)\otimes\Gamma_j
\;+\;
\left(m+\sum_{j=1}^3\frac{1}{2}(V_j+V_j^*)\right)\otimes\Gamma_4
\;,
$$
where $V_j$ are the shift operators on $\ell^2(\ZM^3)$ in the three spacial directions and $m\in\RM$ is a mass term. Due to the anti-commutation relations the Hamiltonian has a CHS $\RCH=\Gamma_5$ and an odd TRS $\STR=\imath\Gamma_2$. Thus $H$ also has an even PHS and is in Class DIII, $j=3$. As such, it has a $\ZM$-invariant. Now diagonalizing $W^*\RCH W=\one\otimes\sigma_3$, the Hamiltonian is of the form $W^*HW=\binom{0\;\;A}{A^*\;0}$ and the integer invariant is $\Ind(EAE+\one-E)$. It can be checked \cite{RSFL} that the values are $1$ for $|m|\in(1,3)$ and $-2$ for $m\in(-1,1)$, and $0$ otherwise. At $m=-3,-1,1,3$ the gap is closed. If one now adds a term to the Hamiltonian that breaks the CHS, but keeps the  odd TRS intact ({\it e.g.} the perturbation $\frac{\lambda}{2\imath}(V_1-V_1^*)\otimes\Gamma_5$ for some $\lambda\in\RM$), the model falls into Class AII and therefore has a $\ZM_2$-invariant in dimension $d=3$. If $\Ind(EAE+\one-E)$ is odd (as for $1<|m|<3$) this $\ZM_2$-invariant given by $\dim(\Ker(EHE))\,\mbox{mod}\,2$ is non-trivial.

\section{Index pairings of $KR$-groups with $KR$-cycles}
\label{sec-NCG}

In this short section, it will be sketched how the functional analytic results of Section~\ref{sec-indsym} can be integrated in the realm of non-commutative geometry \cite{Con,Con2,GVF}. For this purpose, one first of all needs a C$^*$-algebra $\Aa$ with an anti-linear involutive $*$-automorphism $\tau:\Aa\to\Aa$. The notation $\tau(A)=\overline{A}$ will then be used. For sake of concreteness, one may suppose that $\Aa$ is given by a concrete subset of operators on a Hilbert space $\Hh$ with  complex conjugation $\Cc$, and that $\tau(A)=\Cc A\Cc$ for $A\in \Aa$. Inside of $\Aa$ lies a pre-C$^*$-algebra $\Aa_0$. Associated to this data $KR$-cycles and $KR$-groups can be defined (see below) and their pairing can be calculated in an abstract manner as a Kasparov product \cite{Kas}. This leads to the concrete index pairings given by the Fredholm operators in Section~\ref{sec-IndPair}. These operators have numerical invariants $\Ind$ and $\Ind_2$ taking values in $\ZM$, $2\,\ZM$ or $\ZM_2$ which are, by the results of Section~\ref{sec-indsym}, indeed independent of the choice of representative of the $KR$-group.

\vspace{.2cm}

The results of Section~\ref{sec-overview} then appear as a concrete realization of the abstract theory to the C$^*$-algebraic theory of aperiodic media \cite{Bel,BES}. Here the C$^*$-algebra $\Aa$ is the algebra of covariant observables constructed as a reduced crossed product, tensorized with finite matrix algebras. It can be seen as a concrete set of operators on the Hilbert space $\Hh=\ell^2(\ZM^d)\otimes\CM^N$. Moreover, $\Aa_0$ is the set of covariant observables of finite range. The Fermi projection specifies a $KR$-group element and the Dirac operator provides a $KR$-cycle.

\subsection{$KR$-cycles}
\label{sec-KRcycles}

In this section, it is assumed that $\Aa$ is a unital C$^*$-algebra of operators on a Hilbert space with complex conjugation $\Cc$. According to \cite[Definition 9.18]{GVF} a (reduced) $KR^i$-cycle for $(\Aa_0,\tau)$ and $i=0,\ldots,7$ is given by a self-adjoint (Dirac) operator $D$ on $\Hh$ with compact resolvent and bounded commutators $[A,D]$ for $A\in\Aa_0$ as well as a symmetry operator $\widetilde{\Sigma}$ and, for even $i$ another even symmetry operator $\Omega$ (called grading) such that $\Omega^*\, D\,\Omega=-D$ and
\begin{center}
\begin{tabular}{|c|c|c|c|c|c|c|c|c|}
\hline
$i$ & $0$ & $1$ & $2$ &  $3$ &  $4$ &  $5$ &  $6$ &  $7$ 
\\
\hline
$\widetilde{\Sigma}^2$ & $\one$ & $\one$ & $-\one$ & $-\one$ & $-\one$ & $-\one$ & $\one$ & $\one$ 
\\
\hline
$\widetilde{\Sigma}^*\,\overline{D}\,\widetilde{\Sigma}$ & $D$ & $-D$ & $D$ & $D$ & $D$ & $-D$ & $D$ & $D$ 
\\
\hline
$\Omega\,\widetilde{\Sigma}\,\Omega$ & $\widetilde{\Sigma}$ & & $-\widetilde{\Sigma}$ & & $\widetilde{\Sigma}$ & & $-\widetilde{\Sigma}$ & 
\\
\hline
\end{tabular}
\end{center}
These reduced $KR$-cycles are equivalent to Kasparov's $KR$-cycles \cite{Kas}, see \cite[Thm. 9.19]{GVF}. The above is verbatim the definition of \cite{GVF} except for the slight difference that in \cite{GVF} the anti-linear operator $\widetilde{\Sigma}\,\Cc$ appears instead of $\widetilde{\Sigma}$, but this is just a rewriting to match our prior notations. Let us also note that $KR$-cycles are also called spectral triples with real structure \cite{Con2} or $KR$-homology elements. The Dirac operator \eqref{eq-diracop} together with the symmetries \eqref{eq-DiracSymShort} constitutes an example of $KR^i$-cycle provided the following identifications are made: $i=8-d$ and the symmetry operator $\widehat{\Sigma}$ is chosen as $\widetilde{\Sigma}=\widehat{\Sigma}$ for $i=2,6$ and $\widetilde{\Sigma}={\Sigma}$ for all other $i$.

\subsection{Complex $K$-groups}
\label{sec-Kgroups}

As preparation for the $KR$-groups, this section presents the construction of the complex $K$-groups for a given C$^*$-algebra (which may or may not be unital). This does not invoke $\tau$ yet. Inspired by \cite{HL} (see also \cite{VD}), self-adjoint operators $Q=\one-2P$ squaring to the identity will be used for the definition of the $K_0$-group instead of projections $P$. Moreover, $Q$ will be taken as an element in the unitalization $\Aa^+$ as this will readily allow in Proposition~\ref{prop-K1} to connect to the so-called {\it standard picture} for $K$-theory \cite{RLL}. As a vector space one has  $\Aa^+=\Aa\oplus\CM$, but the product is $(A,t)(B,s)=(AB+As+Bt,ts)$ and the adjunction $(A,t)^*=(A^*,\overline{t})$. Equipped with a natural norm, $\Aa^+$ becomes a C$^*$-algebra with unit $\one=(0,1)$. It is part of a split exact sequence of C$^*$-algebras
$$
0 \; \rightarrow  \; \Aa \; \overset{i}{\hookrightarrow} \; \Aa^+  \;  
\overset{\pi}{\rightarrow}
\; \CM \; \rightarrow \; 0\;.
$$
Moreover, there exists an embedding $i':\CM\to\Aa^+$ given by $i'(t)=(0,t)$ which is a right inverse to $\pi$. Then $s=i'\circ \pi:\Aa^+\to\Aa^+$ extracts the scalar part. Set
\begin{equation}
\label{eq-V0def}
V_0(\Aa)
\;=\;
\left\{\left.
Q\in \cup_{n\geq 1}M_{2n}(\Aa^+)\;\right|\;
Q^*\,=\,Q\;,
\;\;
Q^2\,=\,\one\;,
\;\;
s(Q)\sim_0 E_{2n}
\right\}
\;,
\end{equation}
where $M_{2n}(\Aa^+)$ denotes the $2n\times 2n$ matrices over $\Aa^+$, and $s(Q)\sim_0 E_{2n}$ requires the scalar part of $Q$ to be homotopic to $E_{2n}=E_2^{\oplus^n}$ with $E_{2}=\binom{\one\;\;\;\;0\;}{0\;\;-\one}$ in the space of scalar matrices  of adequate size $2n$. On $V_0(\Aa)$ an equivalence relation $\sim_0$ is defined by homotopy (w.r.t. the C$^*$-norm topology) within the self-adjoint unitaries of a given fixed matrix algebra $M_{2n}(\Aa^+)$ and alternatively the requirement
\begin{equation}
\label{eq-equirel}
Q\;\sim_0\;
\begin{pmatrix}
Q & 0 \\
0 & E_2  
\end{pmatrix}
\;\in\;M_{2(n+1)}(\Aa^+)\;,
\qquad
Q\,\in\,M_{2n}(\Aa^+)
\;.
\end{equation}
Then the quotient $V_0(\Aa)\slash\sim_0$ becomes an abelian group with neutral element $0=[E_2]_0$ via
\begin{equation}
\label{eq-semigroup}
[Q]_0\;+\;[Q']_0\;=\;\left[
\begin{pmatrix}
Q & 0 \\ 0 & Q'
\end{pmatrix}
\right]_0
\;.
\end{equation}
This group is by definition the complex $K_0$-group $K_0(\Aa)$. The inverse in $K_0(\Aa)$ is given by $-[Q]_0=[-Q]_0$ and furthermore  $0=[E_{2n}]_0$ for all $n\geq 1$.  The group $K_1(\Aa)$ can be constructed by, moreover, imposing a symmetry (of chiral type in the terminology of Section~\ref{sec-overview}). Hence let us introduce the even symmetry operator $R_{2n}=R_{2}^{\oplus^n}$ with $R_{2}=\binom{0\;\;\one}{\one\;\;0\;}$ extended diagonally $R=\oplus_{n\geq 1}R_{2n}$ to $\cup_{n\geq 1}M_{2n}(\Aa^+)$. Then set
\begin{align*}
V_1(\Aa)
& \;=\;
\left\{\left.
Q\in V_0(\Aa)\;\right|\;
R^*\,Q\,R\,=\,-\,Q
\right\}
\\
& \;=\;
\left\{\left.
Q\in \cup_{n\geq 1}M_{2n}(\Aa^+)\;\right|\;
Q^*\,=\,Q\;,
\;\;
Q^2\,=\,\one\;,
\;\;
R^*\,Q\,R\,=\,-\,Q
\right\}
\;.
\end{align*}
The equality between the two sets holds because $R^*QR=-Q$ implies that $s(Q)$ is homotopic to $E_{2n}$ in $V_1(\Aa)$ by spectral theory. Again an equivalence relation $\sim_1$ on $V_1(\Aa)$ is given by \eqref{eq-equirel} and homotopy, but now within the set $V_1(\Aa)$.  Note  that \eqref{eq-semigroup} conserves the symmetry and that $E_{2n}\in V_1(\Aa)$ still is a unit $0=[E_{2n}]_1$. Hence the quotient $V_1(\Aa)\slash\sim_1$ becomes again a group which is $K_1(\Aa)$ and inverse given by $-[Q]_1=[-E_{2n}QE_{2n}]_1$, see \cite{VD}.

\begin{proposi} 
\label{prop-K1} 
These definitions coincide with those given in the literature, e.g. {\rm \cite{RLL,GVF}}.
\end{proposi}

\noindent {\bf Proof.} Let $\widehat{K}_0(\Aa)$ be the standard picture of  $K_0$ given in \cite[Proposition~4.2.2]{RLL} as
$$
\widehat{K}_0(\Aa)
\;=\;
\left\{\left.[P]_0-[s(P)]_0
\;\right|\;
P=P^2=P^*\in \cup_{n\geq 1}M_{n}(\Aa^+)
\right\}
\;.
$$
Here $s(P)$ is the scalar part of $P\in\Aa^+$ and the difference is formal in the sense of Grothendiek. The equivalence relation is taken on the set of projections in matrix algebras over $\Aa^+$, by homotopy and $P\sim_0\binom{P\;0}{0\;0}$ similar as in \eqref{eq-equirel}. Addition on $\widehat{K}_0(\Aa)$ is defined as in \eqref{eq-semigroup}. Now define $\varphi_0:\widehat{K}_0(\Aa)\to K_0(\Aa)$ by
$$
\varphi_0([P]_0-[s(P)]_0)
\;=\;
\left[
\begin{pmatrix}
\one_n-2P & 0
\\
0 & 2\,s(P)-\one_n
\end{pmatrix}
\right]_0
\;,
\qquad
P\in M_{n}(\Aa^+)
\;.
$$
Now it can be checked that $\varphi_0$ is a group homomorphism with inverse given by 
$$
\varphi_0^{-1}([Q]_0)\;=\;[\chi(Q<0)]_0-[\chi(s(Q)<0)]_0
\;.
$$ 
Next let us connect to the standard way to define $K_1$-groups. Set 
$$
W_1(\Aa)
\;=\;
\left\{\left.
U\in \cup_{n\geq 1}M_{n}(\Aa^+)\;\right|\;
U^{-1}\,=\,U^*
\right\}
\;,
$$
on which the equivalence relation $\sim_1$ defined by homotopy and $[U]_1=[\binom{U\;0}{0\;\one}]_1$. The addition on $W_1(\Aa)/\sim_1$ is again defined by $[U]_1+[U']_1=[U\oplus U']_1$. A standard homotopy argument shows that this is equivalent to $[U]_1+[U']_1=[UU']_1$ and that $[U]_1=[Us(U)^*]_1$ holds. This defines the standard picture $\widehat{K}_1(\Aa)=W_1(\Aa)/\sim_1$ \cite{RLL}. Let $f_{2n}=f_2^{\oplus^n}$ with $f_2=\frac{1}{\sqrt{2}}\binom{\one \;\;\; \one}{\one\;-\one}$. Then $f_2R_2f_2=\binom{\one \;\;\; 0\;}{0\;-\one}$. Hence there is a permutation $g_{2n}$ such that $g_{2n}f_{2n}R_{2n}f_{2n}g_{2n}=\binom{\one_n \;\;\; 0\;}{0\;\;-\one_n}$. Set $O=\oplus_{n\geq 1}f_{2n}g_{2n}$. Then $\widehat{R}=O^*R O=\oplus_{n\geq 1}\binom{\one_n \;\;\; 0\;}{0\;\;-\one_n}$. For $\widehat{Q}=O^*QO$ one has the relations $\widehat{R}^*\widehat{Q}\widehat{R}=-\widehat{Q}$ and  $\widehat{Q}^2=\one$. This implies that there is a unitary $U$ such that 
\begin{equation}
\label{eq-Qform}
\widehat{Q}
\;=\;
\begin{pmatrix}
0 & U \\
U^* & 0
\end{pmatrix}
\;,
\qquad
Q
\;=\;
\frac{1}{2}
\begin{pmatrix}
U+U^* & -U+U^* \\
U-U^* & -U-U^*
\end{pmatrix}
\;.
\end{equation}
Now $\varphi_1:\widehat{K}_1(\Aa)\to K_1(\Aa)$ will be given by $\varphi_1([U]_1)=[O\binom{0 \;\;U}{U^*\;0}O^*]_1$, and it can be checked to be a group isomorphism (note that, in particular, one has the homotopy $s(O\binom{0 \;\;U}{U^*\;0}O^*)\sim E_{2n}$), with inverse $\varphi_1^{-1}([Q]_1)=[\frac{1}{2}\binom{\one}{\one}^*Q\binom{\one}{-\one}]_1$.
\hfill $\Box$

\vspace{.2cm}

Just as $K_1(\Aa)$ was introduced above as classes of projections (or self-adjoints squaring to the identity) satisfying a supplementary symmetry relation, one can also view $K_0(\Aa)$ as classes of even-dimensional unitaries $Q$ satisfying the relations $Q=Q^*$ and $s(Q)\sim_0 E_{2n}$, {\it cf.} \eqref{eq-V0def}.

\subsection{$KR$-groups}
\label{sec-KRgroups}

The groups $KR_j(\Aa,\tau)$ will be defined as the sets of equivalence classes of projections which satisfy the relations specified in the first column of the table in Theorem~\ref{theo-indexlist}. Unfortunately, we were unable to localize a reference where the $KR$-groups are defined in this manner, except for the cases $j=0,1,2$ which are treated in \cite{HL} and a unitary picture which is discussed in the independent and parallel work \cite{BL}. The connections to \cite{Kar,Kas,Sc} will be the object of further investigations (as stated in the Acknowledgements, \cite{VD} and very recently \cite{Kel} already contribute to this point). Again self-adjoints squaring to the identity will be used rather than projections. Hence $\tau$ is extended to $\Aa^+$ as $\tau(A,t)=(\tau(A),\overline{t})$ which is still an anti-linear involutive $*$-automorphism. Also for $a=(A,t)\in\Aa^+$  the notation $\tau(a)=\overline{a}$ is used. Let us start with even $j$ and set
\begin{align*}
& V_0(\Aa,\tau)
\;=\;
\left\{
Q\in V_0(\Aa)\;\left|\;
S_0^*\,\overline{Q}\,S_0\,=\,Q
\;,
\;\;
s(Q)\sim E_{2n}
\right.
\right\}
\;,
& \mbox{(real s.-a.)}\;,
\\
& V_2(\Aa,\tau)
\;=\;
\left\{
Q\in V_0(\Aa)\;\left|\;
S_2^*\,\overline{Q}\,S_2\,=\,-\,Q
\;,
\;\;
s(Q)\sim E_{2n}
\right.
\right\}
\;,
& \mbox{(even symplectic s.-a.)}\;,
\\
& V_4(\Aa,\tau)
\;=\;
\left\{
Q\in V_{0,\mbox{\rm\tiny ev}}(\Aa)\;\left|\;
S_4^*\,\overline{Q}\,S_4\,=\,Q
\;,
\;\;
s(Q)\sim E_{2n}\otimes\one_2
\right.
\right\}
\;,
& \mbox{(quaternionic s.-a.)}\;,
\\
& V_6(\Aa,\tau)
\;=\;
\left\{
Q\in V_0(\Aa)\;\left|\;
S_6^*\,\overline{Q}\,S_6\,=\,-\,Q
\;,
\;\;
s(Q)\sim E_{2n}
\right.
\right\}
\;,
& \mbox{(odd symplectic s.-a.)}\;,
\end{align*}
where s.-a. stands for self-adjoint, the homotopies $s(Q)\sim E_{2n}$ and $s(Q)\sim E_{4n}$ have to respect the required symmetries, and in $V_{0,\mbox{\rm\tiny ev}}(\Aa)$ matrices have doubled dimension 
$$
V_{0,\mbox{\rm\tiny ev}}(\Aa)
\;=\;
\left\{\left.
Q\in \cup_{n\geq 1}M_{4n}(\Aa^+)\;\right|\;
Q^*\,=\,Q\;,
\;\;
Q^2\,=\,\one\;,
\;\;s(Q)\sim E_{2n}\otimes\one_2
\right\}
\;,
$$
and finally the symmetry operators are given by $S_0=\one$ and
$$
S_2
\;=\;\cup_{n\geq 1}
\begin{pmatrix}
0 & \one \\ \one & 0  \end{pmatrix}^{\oplus^n} 
\;,
\qquad
S_4
\;=\;\cup_{n\geq 1}
\begin{pmatrix}
\Sigma_2 & 0 \\ 0 & \Sigma_2  \end{pmatrix}^{\oplus^n} 
\;,
\qquad
S_6
\;=\;\cup_{n\geq 1}
\begin{pmatrix}
0 & - \one \\ \one & 0 \end{pmatrix}^{\oplus^n} 
\;,
$$
where $\Sigma_2=\binom{0\;-\one}{\one\;\;0}$. Above the terminology symplectic self-adjoint for a symmetry of type $S^*\overline{Q}S=-Q$ reflects that the associated  projection $P=\frac{1}{2}(Q+\one)$ has a symplectic symmetry $S^*\overline{P}S=\one-P$, and this (symplectic) Lagrangian  symmetry can be either even or odd pending on the sign of $S^2$. In the terminology of Section~\ref{sec-overview} these symmetries are also called even or odd particle-hole symmetries, while the real and quaternionic symmetries are called even and odd time-reversal symmetry. 

\vspace{.2cm}

Next let us note that $E_2\in V_{j}(\Aa,\tau)$ for $j=0,2,6$ and $E_2\otimes \one_2\in  V_{4}(\Aa,\tau)$.  On each set $V_{2i}(\Aa,\tau)$ an equivalence relation $\sim$ is defined by homotopy and \eqref{eq-equirel}.  As the symmetries of $V_{2i}(\Aa,\tau)$ are conserved under the addition \eqref{eq-semigroup}, it provides a semi-group structure on the quotient $V_{2i}(\Aa,\tau)\slash\sim$ with neutral element $E_2$ for $j=0,2,6$ and $E_2\otimes \one_2$ for $j=4$. The associated Grothendiek group is by definition $KR_{2i}(\Aa,\tau)$. Furthermore let us stress that while the same symbol $\sim$ is used, homotopies are always only constructed within $V_{2i}(\Aa,\tau)$ so that, in particular, it is {\it wrong} to conclude that $KR_{2i}(\Aa,\tau)$ is a subgroup of $K_0(\Aa)=V_0(\Aa)\slash\sim_0$.

\vspace{.2cm}

For odd $j=2i+1$, one now imposes conditions in $V_1(\Aa)$ which correspond to the rows in the table in Theorem~\ref{theo-indexlist} with same $j$, in a manner explained further below. Again we need $V_1(\Aa)$ with doubled dimensions:
$$
V_{1,\mbox{\rm\tiny ev}}(\Aa)
\;=\;
\left\{\left.
Q\in V_{0,\mbox{\rm\tiny ev}}(\Aa)\;\right|\;
R_{\mbox{\rm\tiny ev}}^*\,Q\,R_{\mbox{\rm\tiny ev}}\,=\,-\,Q
\right\}
\;,
$$
where $R_{\mbox{\rm\tiny ev}}=\oplus_{n\geq 1}(R_{2}\otimes\one_2)^{\oplus^n}$.
\begin{align*}
& V_1(\Aa,\tau)
\;=\;
\left\{
Q\in V_1(\Aa)\;\left|\;
S_1^*\,\overline{Q}\,S_1\,=\,Q
\;,
\;\;
s(Q)\sim E_{2n}
\right.
\right\}
\;,
& \mbox{(real c. s.-a.)}\;,
\\
& V_3(\Aa,\tau)
\;=\;
\left\{
Q\in V_{1,\mbox{\rm\tiny ev}}(\Aa)\;\left|\;
S_3^*\,\overline{Q}\,S_3\,=\,-\,Q
\,,
\,
s(Q)\sim E_{2n}\otimes\one_2
\right.
\right\}
\,,
& \mbox{(even symplectic c. s.-a.)}\;,
\\
& V_5(\Aa,\tau)
\;=\;
\left\{
Q\in V_{1,\mbox{\rm\tiny ev}}(\Aa)\;\left|\;
S_5^*\,\overline{Q}\,S_5\,=\,Q
\;,
\;\;
s(Q)\sim E_{2n}\otimes\one_2
\right.
\right\}
\;,
& \mbox{(quaternionic c. s.-a.)}\;,
\\
& V_7(\Aa,\tau)
\;=\;
\left\{
Q\in V_1(\Aa)\;\left|\;
S_7^*\,\overline{Q}\,S_7\,=\,-\,Q
\;,
\;\;
s(Q)\sim E_{2n}
\right.
\right\}
\;,
& \mbox{(odd symplectic c. s.-a.)}\;,
\end{align*}
where c. s.-a. stands for chiral self-adjoints, $S_1=\one$ and  
$$
S_3
\;=\;\cup_{n\geq 1}
\begin{pmatrix}
0 & -\Sigma_2 \\ \Sigma_2 & 0  \end{pmatrix}^{\oplus^n} 
\;,
\qquad
S_5
\;=\;\cup_{n\geq 1}
\begin{pmatrix}
\Sigma_2 & 0 \\ 0 & \Sigma_2  \end{pmatrix}^{\oplus^n} 
\;,
\qquad
S_7
\;=\;\cup_{n\geq 1}
\begin{pmatrix}
0 & - \one \\ \one & 0 \end{pmatrix}^{\oplus^n} \,.
$$
Again $E_2\in V_{j}(\Aa,\tau)$ for $j=1,7$ and $E_2\otimes \one_2\in  V_{j}(\Aa,\tau)$ for $j=3,5$. Furthermore, \eqref{eq-equirel} defines an equivalence relation $\sim$ on $V_j(\Aa,\tau)$ and \eqref{eq-semigroup} a semi-group structure on the quotient $V_j(\Aa,\tau)\slash\sim$ with neutral elements. The associated Grothendiek groups are now by definition $KR_j(\Aa,\tau)$ for odd $j$.

\vspace{.2cm}

Let us point out that the odd $V_{2n+1}(\Aa,\tau)$ are obtained from $V_{2n}(\Aa,\tau)$ by imposing the chirality constraint while keeping $S_{2n}=S_{2n+1}$, except for the pair $j=2,3$ because in $j=3$ one requires the chiral symmetry $R$ to anti-commute with the even symplectic symmetry, and this requires to pass to $2\times 2$ matrices so that $S_2$ and $S_3$ are different (but still both commute to $\one$ and the symmetry relation is in both cases $S^*\overline{Q}S=-Q$).

\vspace{.2cm}

Instead of the commuting symmetry operators $S$ and $\widehat{S}$ of Theorem~\ref{theo-indexlist} we rather used $S$ and $R=S\widehat{S}$. This immediately leads to the formulas above in the cases $j=1,5$. For $j=3,7$, however, a Cayley transform is needed as in Section~\ref{sec-CGLieG} (or the proof of Proposition~\ref{prop-FU}) in order to transform $S\widehat{S}$ with $(S\widehat{S})^2=-\one$ to the symmetry operator $R$ used in the definition of $V_1(\Aa)$ above. This leads to anti-commuting symmetries for $j=3,7$: 
$$
RS_1\,=\,S_1R\;,
\quad
R_{\mbox{\rm\tiny ev}} S_3\,=\,-S_3R_{\mbox{\rm\tiny ev}}\;,
\quad
R_{\mbox{\rm\tiny ev}} S_5\,=\,S_5R_{\mbox{\rm\tiny ev}}\;,
\quad
RS_7\,=\,-S_7R
\;.
$$

\vspace{.2cm}

Just as $K_1(\Aa)$, it is also possible to view the odd $KR$-groups as classes of unitaries. For this set $\Sigma=\cup_{n\geq 1}\Sigma_2^{\oplus^n}$. By reducing out the definitions for $V_{2i+1}(\Aa,\tau)$ one finds that with
\begin{align*}
& W_1(\Aa,\tau)
\;=\;
\left\{
U\in W_1(\Aa)\;\left|\;
\overline{U}\,=\,U
\right.
\right\}
\;,
& \mbox{(real unitary)}\;,
\\
& W_3(\Aa,\tau)
\;=\;
\left\{
U\in W_{1,\mbox{\rm\tiny ev}}(\Aa)\;\left|\;
\Sigma^*\,\overline{U}\,\Sigma\,=\,U^*
\right.
\right\}
\;,
& \mbox{(odd-symmetric unitary)}\;,
\\
& W_5(\Aa,\tau)
\;=\;
\left\{
U\in W_{1,\mbox{\rm\tiny ev}}(\Aa)\;\left|\;
\Sigma^*\,\overline{U}\,\Sigma\,=\,U
\right.
\right\}
\;,
& \mbox{(quaternionic unitary)}\;,
\\
& W_7(\Aa,\tau)
\;=\;
\left\{
U\in W_1(\Aa)\;\left|\;
\overline{U}\,=\,U^*
\right.
\right\}
\;,
& \mbox{(symmetric unitary)}\;,
\end{align*}
where
$$
W_{1,\mbox{\rm\tiny ev}}(\Aa)
\;=\;
\left\{\left.
U\in \cup_{n\geq 1}M_{2n}(\Aa^+)\;\right|\;
U^{-1}\,=\,U^*
\right\}
\;.
$$
With these definitions, one can check $KR_{2i+1}(\Aa,\tau)=W_{2i+1}(\Aa,\tau)\slash \sim_1$ in a similar manner as in Proposition~\ref{prop-K1} using that  $Q\in V_1(\Aa)$ is given in terms of a unitary by \eqref{eq-Qform}. One could also express the even $KR$-groups by unitaries, but this is not written out here (see \cite{BL}).

\vspace{.2cm}

\appendix

\section{Symmetry operators for the Clifford groups}
\label{app-Cd}

The definition \eqref{eq-diracop} of the Dirac operator involves the $\Gamma$-matrices and their construction as well as their symmetries are reviewed in this appendix. The complex Clifford algebra $C_d$ for $d\in \NM$ is the complex algebra generated by $\gamma_1,\ldots,\gamma_d$ satisfying $\gamma_n\gamma_m+\gamma_m\gamma_n=2\,\delta_{n,m}$. It will be convenient to first look at the finite subgroup of the multiplicative group of $C_d$ generated by the $\gamma_n$'s, called the Clifford group (Section~\ref{sec-CG}). Then irreducible representations of this group are constructed (Section~\ref{sec-CGrep}) and symmetry operators in these representations are studied (Section~\ref{sec-CGrepSym}). Finally Lie algebras and groups naturally associated to these representations will be constructed (Section~\ref{sec-CGLie} and \ref{sec-CGLieG}). Useful references containing a lot of the material below are \cite{ABS,Kar,LM,Por,GVF}, but we could not locate all of the facts and felt that a self-contained description adapted to the needs in the main text is worth while including as an appendix. Furthermore, Sections~\ref{sec-CGLie} and \ref{sec-CGLieG} exhibit a close connection of the Clifford generators to Weyl's classical matrix groups which complements the existing literature \cite{Por}.

\subsection{Clifford groups}
\label{sec-CG}

The Clifford group $F_d$ with identity $1$ is generated by $\gamma_1,\ldots,\gamma_d$ and an extra element $-1$ commuting with all $\gamma_n$, such that relations $(-1)^2=\gamma_n^2=1$ and $\gamma_n\gamma_m=(-1)\gamma_m\gamma_n$ for $n\not=m$ hold. Clearly $F_{d-1}\subset F_d$ is a subgroup. Actually, one has $F_d=F_{d-1}\cup \gamma_d F_{d-1}$. This implies that $F_d$ has $2^{d+1}$ elements. Three involutive automorphisms $\alpha,\beta,\widehat{\alpha}:F_d\to F_d$ will play a crucial role below. They are defined by, for $n=1,\ldots,d$,
$$
\alpha(\gamma_n)
\;=\;
\kappa\,(-1)^{n+1}\,
\gamma_{n}
\;,
\qquad
\beta(\gamma_n)
\;=\;
-\,\gamma_n
\;,
\qquad
\widehat{\alpha}(\gamma_n)
\;=\;
\kappa\,(-1)^n\,
\gamma_{n}
\;,
$$
together with $\alpha(\pm1)=\pm 1$, $\beta(\pm1)=\pm 1$ and $\widehat{ \alpha}(\pm1)=\pm 1$. Here $\kappa\in\{-1,1\}$ is given by  $\kappa=(-1)^{\lfloor \frac{d}{2}\rfloor}$. Note that all these objects depend on $d$, but we place an index only when it is unavoidable. Clearly $\alpha\circ\beta=\beta\circ\alpha=\widehat{\alpha}$. The peculiar choice of the sign $\kappa$ is crucial because it insures by Proposition~\ref{prop-adjoinable} below that $\alpha$ is always adjoinable (one also says $\alpha$ is inner), namely that there exists a $\sigma\in F_d$ with $\alpha(\gamma)=\sigma\,\gamma\,\sigma^{-1}$ for all $\gamma\in F_d$. The involutions $\beta$ and $\widehat{\alpha}$ turn out to be adjoinable only for even $d$. In many prior works starting with \cite{ABS}, the involution $\beta$ played a prominent role and is called the main involution. Let us introduce
$$
\sigma
\;=\;
\gamma_{2}\gamma_4\gamma_6\cdots\gamma_{2\,\lfloor \frac{d}{2}\rfloor}
\;,
\qquad
\omega
\;=\;
\gamma_{1}\gamma_2\gamma_3\cdots\gamma_{d}
\;,
\qquad
\widehat{\sigma}
\;=\;
\gamma_{1}\gamma_3\gamma_5\cdots\gamma_{2\,\lfloor \frac{d}{2}\rfloor+1}
\;.
$$
Often $\omega$ is called the volume element. One checks
$$
\sigma^2
\;=\;
\left\{
\begin{array}{cc}
1\;, & d=1\;,
\\
-\,\kappa\;, & d\not = 1\;,
\end{array}
\right.
\qquad
\omega^2
\;=\;
(-1)^{d-1}
\;,
\qquad
\widehat{\sigma}^2
\;=\;
(-1)^{\lfloor \frac{d-1}{2}\rfloor}
\;,
\qquad
\omega\,\sigma\;=\;(\kappa)^d\,\widehat{\sigma}
\;.
$$
Using $\omega^{-1}=\gamma_{d}\gamma_{d-1}\cdots\gamma_{1}$ one readily checks
$$
\omega\,\gamma_n\,\omega^{-1}\;=\;(-1)^{d-1}\,\gamma_n
\;,
\qquad
n=1,\ldots,d
\;.
$$
Thus $\omega$ is central for odd $d$, and actually the center is $\{1,-1,\omega,(-1)\omega\}$ for odd $d$. On the other hand,  for even $d$ the center is $\{1,-1\}$  and one can decompose $F_d=F_{d,+}\cup F_{d,-}$ into elements having an even or odd number of $\gamma_n$'s, that is, $F_{d,\pm}=\{\gamma\in F_d\,|\,\omega\,\gamma\,\omega^{-1}=(\pm 1) \gamma\}$.

\begin{proposi} 
\label{prop-adjoinable} 
The automorphism $\alpha$ is adjoinable and
\begin{equation}
\label{eq-alphasigma}
\alpha(\gamma)\;=\;\sigma\,\gamma\,\sigma^{-1}\;,
\qquad
\gamma\in F_d
\;.
\end{equation}
The automorphisms $\beta$ and $\widehat{\alpha}$ are not adjoinable for odd $d$, and for even $d$ one has
$$
\beta(\gamma)\;=\;\omega\,\gamma\,\omega^{-1}\;,
\qquad
\widehat{\alpha}(\gamma)\;=\;\widehat{\sigma}\,\gamma\,\widehat{\sigma}^{-1}\;,
\qquad
\gamma\in F_d
\;.
$$
\end{proposi}

\noindent {\bf Proof.} For the first claim, it is sufficient to check that $\sigma\,\gamma_n\,\sigma^{-1}=\kappa (-1)^{d+1}\gamma_n$. This is a matter of patience. Let now $d$ be odd and let us first show that $\beta$ is not adjoinable. Suppose that there is an $\eta\in F_d$ with $\eta\,\gamma_n\,\eta^{-1}=-\gamma_n$ for all $n=1,\ldots,d$. Then $\eta\, \omega\, \eta^{-1}=(-1)^d\omega=(-1)\omega$, which is in contradiction to the fact that $\omega$ is central for odd $d$. For $\widehat{\alpha}$ one can argue similarly, or use that its adjoinability would imply the adjoinability of $\beta$. The formulas for even $d$ can again be readily checked.
\hfill $\Box$

\vspace{.2cm}

Let us point out that $\sigma$ is not uniquely specified by \eqref{eq-alphasigma}. In fact, for any $\eta$ in the center of $F_d$, one has $\alpha(\gamma)=(\eta\,\sigma)\,\gamma\,(\eta\,\sigma)^{-1} $. For even $d$, this only allows for a sign change of $\sigma$, but  for odd $d$ one can choose $\eta=\omega$ so that $\alpha(\gamma)\;=\;\widehat{\sigma}\,\gamma\,\widehat{\sigma}^{-1}$. Indeed, $\omega\,\sigma$ is either equal to $\widehat{\sigma}$ or to $(-1)\widehat{\sigma}$. Furthermore let us point out that $\omega$ ad $\widehat{\sigma}$ are unique up to a sign for even $d$.

\subsection{Representations of the Clifford group}
\label{sec-CGrep}

Here the task is to construct a particular unitary, irreducible, complex representation $\pi_d:F_d\to\CM^{d'\times d'}$ such that $\pi_d(\gamma_{2n+1})$ are real and $\pi_d(\gamma_{2n})$ are purely imaginary. This alternation is feasible in an irreducible complex representation and it turns out that the dimension of the representation is $d'= 2^{\lfloor \frac{d}{2}\rfloor}$, except for the special case of $d=1$ for which $1'=2$. The dimension of a real representation, namely by real matrices $\pi_d(\gamma_{n})$, would be larger. This construction is done iteratively using the $2\times 2$ Pauli matrices
\begin{equation}
\label{eq-Pauli} 
\sigma_1\;  =\; \begin{pmatrix}0 & 1\\1 & 0
\end{pmatrix}
 \, , 
\qquad
\sigma_2 \,= \,
\begin{pmatrix} 0 & -\imath \\ \imath & 0
\end{pmatrix}\, , 
\qquad
\sigma_3 \,= \,\begin{pmatrix}1 & 0 \\ 0 & -1
 \end{pmatrix} \,  .
\end{equation}
Also let $\one_n\in\CM^{n\times n}$ denote the identity. Then $\pi_d(-1)=-\one_{d'}$ and
\begin{equation}
\label{eq-CliffDef}
\pi_d(\gamma_{n})\;=\;\sigma_3\otimes\pi_{d-2}(\gamma_{n-2})
\;,
\qquad
\pi_d(\gamma_1)\;=\;\sigma_1\otimes\one_{\frac{d'}{2}} \;,
\;\;\;
\pi_d(\gamma_2)\;=\;\sigma_2\otimes\one_{\frac{d'}{2}}
\;.
\end{equation}
In the following, we will use the short notation $\Gamma_n=\pi_d(\gamma_n)$. Note that for even $d$, the representations $\pi_d$ are given by the restriction of $\pi_{d+1}$. For sake of concreteness, let us write out the generators of the first few representations explicitly:
\begin{align*}
F_1 \, &: \;\; \Gamma_1 = \sigma_1 
\\
F_2 \mbox{ and } F_3 \, &:\;\; \Gamma_1 = \sigma_1  \, , \;\; \Gamma_2 = \sigma_2 , \;\; \Gamma_3 = \sigma_3
\\
F_4 \mbox{ and } F_5 \, &:\;\; \Gamma_1 = \sigma_1  \otimes \one_2 \, , \;\; \Gamma_2 = \sigma_2 \otimes \one_2, \;\; \Gamma_3 = \sigma_3 \otimes \sigma_1  \,, \;\; \Gamma_4 = \sigma_3 \otimes \sigma_2\,, \;\; \Gamma_5 = \sigma_3 \otimes \sigma_3 
\\
F_6 \mbox{ and } F_7 \, &:\;\; \Gamma_1 = \sigma_1  \otimes \one_2\otimes \one_2 \, , \;\; \Gamma_2 = \sigma_2 \otimes \one_2\otimes \one_2, \;\; \Gamma_3 = \sigma_3 \otimes \sigma_1  \otimes \one_2\,, \;\; 
 \\
 &  \;\; \;\;\;\Gamma_4 = \sigma_3 \otimes \sigma_2\otimes \one_2\,, \;\; \Gamma_5 = \sigma_3 \otimes \sigma_3\otimes \sigma_1 \,, \;\; \Gamma_6 = \sigma_3 \otimes \sigma_3\otimes \sigma_2, \;\; \Gamma_7 = \sigma_3 \otimes \sigma_3\otimes \sigma_3
  \,.
\end{align*}

The above representation $\pi_d$ naturally extends to a complex irreducible representation of the Clifford algebra $C_d$. Clearly, $\pi_d(C_d)$ is subalgebra of $\pi_{d+1}(C_{d+1})$ for even $d$. Furthermore,  $\one_2\otimes \pi_{d-1}(C_{d-1})$ is a subalgebra of $\pi_d(C_d)$, still for even $d$. The representation~\eqref{eq-CliffDef}  is not unique as unitary, irreducible representation with real $\Gamma_{2n+1}$ and purely imaginary $\Gamma_{2n}$. There are always unitarily equivalent representations, but arguments similar as in the proof of Proposition~\ref{prop:unitaries-commute-1} allow to reduce to the above for even $d$. On the other hand, it is well-known that for odd $d$ there are two inequivalent representations (a second one is obtained by changing the sign of just one of the $\Gamma$'s, say $\Gamma_d$; this changes the sign of $\Gamma_1\cdots\Gamma_d$ which is, moreover, proportional to the identity). One particular unitarily equivalent representation is obtained by exchanging the roles of $\sigma_1$ and $\sigma_3$, and simultaneously adding a sign to each $\sigma_2$. The real unitary basis change is 
$$
f
\;=\;
\frac{1}{\sqrt{2}}\,
\begin{pmatrix}
1 & 1 \\ 1 & -1
\end{pmatrix}
\;=\;f^*\;=\;f^{-1}
\;,
$$
which satisfies $f^*\sigma_1f=\sigma_3$ and $f^*\sigma_2 f=-\sigma_2$ and can be extended as $f^{\otimes^{\lfloor\frac{d}{2}\rfloor}}$ to the representation of $F_d$. It is also possible to cyclically exchange the roles of the Pauli matrices using the Cayley transform
\begin{equation}
\label{eq-Cayley}
c
\;=\;
\frac{1}{\sqrt{2}}\,
\begin{pmatrix}
1 & -\imath \\ 1 & \imath
\end{pmatrix}
\;.
\end{equation}
It satisfies $c^3=e^{-\frac{\imath\pi}{4}}\,\one$ as well as
\begin{equation}
\label{eq-CayleyShift}
c\,\sigma_1\,c^*\;=\;\sigma_2
\;,
\qquad
c\,\sigma_2\,c^*\;=\;\sigma_3
\;,
\qquad
c\,\sigma_3\,c^*\;=\;\sigma_1
\;.
\end{equation}
Conjugating with $c^{\otimes^{{\lfloor\frac{d}{2}\rfloor}}}$  leads to a unitarily equivalent representation $\Gamma'_n$, $n=1,\ldots,d$. This representation has real $\Gamma'_{2n+1}$ and imaginary $\Gamma'_{2n}$, except for $\Gamma'_d$ in the case of odd $d$ which is always real.



\subsection{Symmetry operators in the representation of the Clifford group}
\label{sec-CGrepSym}

Next let us consider the representations of the elements $\sigma$, $\omega$ and $\widehat{\sigma}$ introduced in Section~\ref{sec-CG}. They will be decorated by phase factors so that they become symmetry operators in the sense of Definition~\ref{def-SymmetryOp}:
$$
\Sigma
\;=\;
\imath^{\lfloor\frac{d}{2}\rfloor}\,\pi_d(\sigma)
\;,
\qquad
\Omega
\;=\;
(-\imath)^{\lfloor\frac{d}{2}\rfloor}\;\pi_d(\omega)
\;,
\qquad
\widehat{\Sigma}
\;=\;
(\kappa)^d\;\pi_d(\widehat{\sigma})
\;.
$$
Here the imaginary factors are important in order to produce real matrices, but the signs are merely added to produce nice formulas in \eqref{eq-GammaDef}, \eqref{eq-SigmaDef} and \eqref{eq-SigmaHatDef}. Again it may be better to add upper indices $\Sigma^d$, $\Omega^d$ and $\widehat{\Sigma}^d$, but for sake of notational simplicity this is suppressed as long as there is no ambiguity. The added factors are chosen such that $\Sigma$, $\Omega$ and $\widehat{\Sigma}$ are real and one has
\begin{equation} 
\label{eq-GammaDef}
\widehat{\Sigma}\;=\;\Omega\,\Sigma
\;,
\qquad
\Omega
\;=\;
\left\{
\begin{array}{cc}
\sigma_3^{\otimes^{\frac{d}{2}}}
\;, &
d\;\mbox{\rm even}\;,
\\
\one\;, &
d\;\mbox{\rm odd}\;.
\end{array}
\right.
\end{equation}
Note that $\Omega=\Gamma_{d+1}$ for even $d$. Let us also point out that there is an alternative construction of $\Sigma$ which, using the supplementary upper index on $\Sigma^d$, is iteratively given by
\begin{equation} 
\label{eq-SigmaDef}
\Sigma^d
\;=\;
-\,\Sigma^{d-4}\otimes \sigma_1\otimes\imath\sigma_2
\;,
\qquad
\Sigma^0=\Sigma^1=1\;,
\;\;\;\Sigma^2=\Sigma^3=\imath\sigma_2
\;.
\end{equation}
Furthermore, one has for some sign $\eta$
\begin{equation} 
\label{eq-SigmaHatDef}
\widehat{\Sigma}^d
\;=\;
\eta\;\Sigma^{d-1}\otimes \sigma_1
\;,
\qquad
d\;\mbox{\rm even}
\;.
\end{equation}
For sake of concreteness, let us write out these symmetry operators for low dimensions:

\vspace{.1cm}

\begin{center}
\begin{tabular}{|c||c|c|c|c|c|c|c|c|}
\hline
$d$  & $1$ & $2$ & $3$ & $4$ & $5$ & $6$ & $7$ & $8$ 
\\
\hline
\hline
$\!\Omega\!$  &  & $\!\sigma_3\!$ &  & $\!\sigma_3\otimes \sigma_3\!$ &  & $\!\sigma_3\otimes \sigma_3\otimes \sigma_3\!$ &  & $\!\sigma_3\otimes \sigma_3\otimes \sigma_3\otimes \sigma_3\!$ 
\\
\hline
$\!\Sigma\!$  & $\!1\!$ & $\!\imath\sigma_2\!$ & $\!\imath\sigma_2\!$ & $\!\!-\sigma_1  \otimes \imath \sigma_2\!$ & $\!\!-\sigma_1  \otimes \imath \sigma_2\!$ & $\!\!-\imath\sigma_2  \otimes \sigma_1\otimes \imath\sigma_2\!$ & $\!\!-\imath \sigma_2 \otimes \sigma_1 \otimes \imath \sigma_2\!$ & $\!\sigma_1 \otimes \imath \sigma_2\otimes \sigma_1 \otimes \imath \sigma_2\!$ 
\\
\hline
$\!\widehat{\Sigma}\!$  &  & $\!\sigma_1\!$ &  & $\!\imath\sigma_2  \otimes \sigma_1\!$ &  & $\!\!-\sigma_1  \otimes \imath\sigma_2\otimes \sigma_1\!$ &  & $\!\imath \sigma_2 \otimes \sigma_1\otimes \imath \sigma_2 \otimes \sigma_1\!$ 
\\
\hline
\end{tabular}
\end{center}

\vspace{.2cm}

\noindent The main properties of $\Omega$ and $\Sigma$ are recollected in the table below and can be checked with some care and patience. Moreover, these symmetry operators can be uniquely characterized.


\begin{proposi} 
\label{prop-SigmaUnique} 
Consider $\pi_d(C_d)\subset \CM^{d'\times d'}$ with $d'=2^{\lfloor \frac{d}{2}\rfloor}$. Set $\kappa=(-1)^{\lfloor \frac{d}{2}\rfloor}$ as above.

\begin{enumerate}[{\rm (i)}]

\item Up to a sign there is a unique real unitary $\Sigma\in\CM^{d'\times d'}$, given by the construction above, satisfying
\begin{equation}
\label{eq-SigmaRel}
\Sigma^*\,\Gamma_{n}\;\Sigma
\;=\;
-\,\kappa\,(-1)^n\,
\Gamma_{n}
\;,
\qquad
n=1,\ldots,d
\;.
\end{equation}
This can be restated as $\Sigma^*\,\overline{\Gamma_{n}}\;\Sigma
\,=\,\kappa\,\Gamma_{n}$.

\item For even $d$ and up to a sign, there is a unique real unitary $\Omega\in\CM^{d'\times d'}$, given by {\rm \eqref{eq-GammaDef}},  satisfying 
\begin{equation}
\label{eq-GammaRel}
\Omega^*\,\Gamma_n\,\Omega
\;=\;
-\,\Gamma_n
\;,
\qquad
n=1,\ldots,d
\;. 
\end{equation}

\item For even $d$ and up to a sign, there is a unique real unitary $\widehat{\Sigma}$ characterized by 
\begin{equation}
\label{eq-SigmaHatRel}
\widehat{\Sigma}^*\,\Gamma_{n}\;\widehat{\Sigma}
\;=\;
\kappa\,(-1)^n\,
\Gamma_{n}
\;,
\qquad
n=1,\ldots,d
\;,
\end{equation}
which can be restated as $\widehat{\Sigma}^*\,\overline{\Gamma_{n}}\;\widehat{\Sigma}=-\,\kappa\,\Gamma_{n}$. For odd $d$ there  exists no $\widehat{\Sigma}$ satisfying {\rm \eqref{eq-SigmaHatRel}}. 

\item The following relations hold:
\begin{center}

\begin{tabular}{|c||c|c|c|c|c|c|c|c|}
\hline
$d\,\mbox{\rm mod}\,8$ & $1$ & $2$ & $3$ & $4$ & $5$ & $6$ & $7$ & $8$
\\
\hline\hline
$\Sigma^2=$ &  $\one$ & $-\one$ & $-\one$ & $-\one$ & $-\one$ & $\one$ & $\one$ & $\one$  
\\
$\kappa=$ &  $1$ & $-1$ & $-1$ & $1$ & $1$ & $-1$ & $-1$ & $1$  
\\
$\Omega^2=$ &  & $\one$ &  & $\one$ &  & $\one$ & & $\one$ 
\\
$\Omega\Sigma=$ &  & $-\Sigma\Omega$ &  & $\Sigma\Omega$ &  & $-\Sigma\Omega$ & & $\Sigma\Omega$ 
\\
\hline
$\widehat{\Sigma}^2=$ &  & $\one$ &  & $-\one$ &  & $-\one$ & & $\one$ 
\\
$\widehat{\Sigma}\Sigma=$ &  & $-\Sigma \widehat{\Sigma}$ &  & $\Sigma \widehat{\Sigma}$ &  & $-\Sigma \widehat{\Sigma}$ & & $\Sigma \widehat{\Sigma}$ 
\\
$\Omega\widehat{\Sigma}=$ &  & $-\widehat{\Sigma}\Omega$ &  & $\widehat{\Sigma}\Omega$ &  & $-\widehat{\Sigma}\Omega$ & & $\widehat{\Sigma}\Omega$ 
\\
[0.1cm]
\hline
\end{tabular}
\end{center}
\end{enumerate}
\end{proposi}

\noindent {\bf Proof.} (i)  For $d=1$ the claim is obvious. Hence let us suppose $d\geq 2$ and first consider the case of even $d$. The proof will be done by induction of $d$ and therefore an upper index $d$ will be added on $\Gamma_j^d$ and $\Sigma^d$ for sake of clarity. Let us start by writing $\Gamma^d_1$, $\Gamma^d_2$ and $\Sigma^d$ in the grading of the first fiber of $\CM^{d'}=\CM^2\otimes\CM^{\frac{d'}{2}}$:
$$
\Gamma^d_1
\;=\;
\begin{pmatrix}
0 & \one \\ \one & 0
\end{pmatrix}
\;,
\qquad
\Gamma^d_2
\;=\;
\begin{pmatrix}
0 & \one \\ -\one & 0
\end{pmatrix}
\;,
\qquad
\Sigma^d
\;=\;
\begin{pmatrix}
a & b \\ c & d
\end{pmatrix}
\;.
$$
The imposed commutation relations can be written as $(\Gamma^d_1)^*\Sigma^d\Gamma^d_1=\kappa\,\Sigma^d$ and $(\Gamma^d_2)^*\Sigma^d\Gamma^d_2=-\kappa\,\Sigma^d$. A short calculation shows that with a real unitary $b=\Sigma'$ acting on $\CM^{\frac{d'}{2}}$ one has
$$
\Sigma^d
\;=\;
\begin{pmatrix}
0 & \Sigma' \\ \kappa\, \Sigma' & 0
\end{pmatrix}
\;=\;
\left\{
\begin{array}{cc}
\sigma_1\otimes \Sigma'\;, &\;\;\;\kappa=1\;,
\\
\imath\sigma_2\otimes \Sigma'\;, &\;\;\;\kappa=-1\;.
\end{array}
\right.
$$
Now for $j\geq 3$ one has $\Gamma^d_j=\sigma_3\otimes\Gamma^{d-2}_{j-2}$ by definition and hence, due to the anti-commutation property in the first factor, the other required relations become 
$$
(\Sigma')^*\,{\Gamma^{d-2}_{j-2}}\;\Sigma'\;=\;\kappa\;(-1)^{j-2}\;\Gamma^{d-2}_{j-2}
\;,
\qquad
j=3,\ldots,d
\;.
$$
But this determines $\Sigma'$ uniquely by induction hypothesis (up to a sign). Hence the uniqueness of $\Sigma^d$ holds for even $d$. For odd $d$, $\Sigma^d=\Sigma^{d-1}$ is also unique and actually does have the right commutation relation with $\Gamma_d^d$. (ii) The proof is left to the reader, as it is similar, but more simple than the above argument. (iii) The proof for $\widehat{\Sigma}$ is identical to (i), except that for odd $d$ the required commutation relation with $\Gamma_d^d$ does not hold. (iv) is a matter of calculation.
\hfill $\Box$

\vspace{.2cm}


The following general result allows to bring the two commuting or anti-commuting symmetry operators into a normal form. It will be applied to the representations of the Clifford group.

\begin{proposi} 
\label{prop:unitaries-commute-1}
Let $\Omega$ and $\Sigma$ be symmetry operators on a Hilbert space with complex conjugation. Suppose that $\spec(\Omega)=\{-1,1\}$ and let $\eta,\tau\in\{-1,1\}$ be such that $\Sigma^2=\eta\one$ and $\Omega\Sigma=\tau\Sigma\Omega$. Then, in each of the following cases, there is a real unitary $O$ with

\begin{enumerate}[{\rm (i)}]

\item $O^* \Omega O =\sigma_3\otimes\one$ and $O^* \Sigma O =\sigma_1\otimes\one$  for $\eta = 1$ and $\tau = -1$.

\item $O^* \Omega O =\sigma_3\otimes\one$ and $O^* \Sigma O =\imath\sigma_2\otimes\one$  for $\eta =- 1$ and $\tau = -1$.

\item $O^* \Omega O =\sigma_3\otimes\one$ and $O^* \Sigma O =\one\otimes\imath\sigma_2$  for $\eta = -1$ and $\tau = 1$.

\item $O^* \Omega O =\sigma_3\otimes\one\otimes\one$ and $O^* \Sigma O =\one\otimes\sigma_3\otimes\one$  for $\eta = 1$ and $\tau = 1$ and provided that $\Omega$ and $\Sigma$ have $4$ common eigenspaces of the same dimension.

\end{enumerate}
\end{proposi}

\noindent {\bf Proof.} The eigenspaces of $\Omega$ are invariant under complex conjugation. Hence it is possible to choose a real basis for both of them. These basis are by hypothesis of equal dimension (because this is part of Definition~\ref{def-Symmetries}) and constitute $O'$ satisfying $(O')^* \Omega O' =\sigma_3\otimes\one$. If now $\tau=-1$, one has $(O')^* \Sigma O'=\binom{0\;\;\;\;a}{\eta a^*\;0}$ with some real unitary $a$. Setting $O=O'\binom{a\;\;0}{0\;\;\one}$ concludes the proof of (i) and (ii). If $\tau=1$, then $(O')^* \Sigma O'=\binom{a\;\;0}{0\;\;b}$ with real unitaries $a$ and $b$. If $\eta=-1$ both of these can be diagonalized to $\imath\sigma_2$ by a real unitary. Indeed, the spectrum of $a$ is $\{\imath,-\imath\}$ and the eigenspaces $\Ee_{-\imath}$ and $\Ee_{\imath}$ are complex conjugates of each other and are, in particular, of same dimension. Hence there is a unitary $V=(\overline{v},v)$ built from the basis $v=(v_1,v_2,\ldots)$ of $\Ee_\imath$ such that $V^*a V=-\imath\,\sigma_3\otimes\one$. Now the Cayley transform $C=c\otimes \one$ with $c$ defined in \eqref{eq-Cayley} leads to a real unitary $VC$ which satisfies $(VC)^* a\, VC=\imath\sigma_2\otimes\one$. Similarly $(WC)^*b\,WC=\imath\sigma_2\otimes\one$ and then $O=O'\binom{ VC \;\;\;\;\; 0}{ \;\;0 \;\;\;\; WC}$ has the desired properties. The proof of (iv) is done by diagonalizing $a$ and $b$ and using the supplementary hypothesis. 
\hfill $\Box$

\begin{proposi} 
\label{prop-GammaSigmaNormal}
The symmetries $\Omega$ and $\Sigma$ defined by {\rm \eqref{eq-GammaDef}} and {\rm \eqref{eq-SigmaDef}} have common eigen\-spaces of same dimension. Hence also item {\rm (iv)} of {\rm Proposition~\ref{prop:unitaries-commute-1}} applies.
\end{proposi}

\noindent {\bf Proof.}
The following argument also allows to construct the basis changes $O$ in Proposition~\ref{prop:unitaries-commute-1} explicitly for all $4$ cases when $\Omega$ and $\Sigma$ are the symmetry operators of the Clifford algebra. Let us introduce the vectors $e_1=\binom{1}{0}$ and $ e_{-1}=\binom{0}{1}$ in $\CM^2$. Furthermore, for a multi-index $i=(i_1,\ldots,i_d)\in\{-1,1\}^d$ let us introduce the complementary index $i^c$ obtained by flipping all signs. Now the vectors $v_i=e_{i_1}\otimes\ldots\otimes e_{e_d}$ with $i$ running through $\{-1,1\}^d$ constitute an orthonormal basis of $\CM^{d'}$. Moreover, this is an eigenbasis of $\Omega$ and $\Omega v_i=i_1\cdots i_d \,v_i$. Now one checks that for some sign $\kappa$
$$
(v_j)^*\,\Sigma \,v_i\;=\;\kappa\;\delta_{j=i^c}
\;,
\qquad
(v_i)^*\,\Sigma \,v_j\;=\;\kappa\;\delta_{j=i^c}
\;.
$$
Consequently $\Sigma$ is block diagonal in this basis with $2\times 2$ blocks given by $\binom{0\;\;1}{1\;\;0}$ and $\binom{\;\;0\;\;-1}{-1\;\;\;0}$. Both blocks have eigenvalues $1$ and $-1$ of equal multiplicity. 
\hfill $\Box$

\vspace{.2cm}

Combining Propositions~\ref{prop:unitaries-commute-1} and \ref{prop-GammaSigmaNormal}, one finds that there is always an adequate orthogonal basis change $O$ that brings the two symmetries in the normal form given in the following table:
\begin{equation}
\label{tab-normal}
\begin{tabular}{|c||c|c|c|c|c|c|c|c|c|c|c|c|}
\hline
$d\,\mbox{\rm mod}\,8$ & $1$ & $2$ & $3$ & $4$ & $5$ & $6$ & $7$ &  $8$
\\
\hline
$\!O^*\Sigma O=\!$ & $\!\sigma_1\otimes\one\!$ &  $\!\imath\sigma_2\otimes\one\!$ & $\!\imath\sigma_2\otimes\one\!$ & $\!\one\otimes\imath\sigma_2\!$ & $\!\imath\sigma_2\otimes\one\!$ & $\!\sigma_1\otimes\one\!$  & $\!\sigma_1\otimes\one\!$ &  $\!\one\otimes\sigma_1\otimes\one\!$ 
\\
$\kappa$ & $1$ & $-1$ & $-1$  & $1$  & $1$ & $-1$ & $-1$ & $1$
\\
$\!O^*\Omega O=\!$ &  & $\!\sigma_3\otimes\one\!$ &  & $\!\sigma_3\otimes\one\!$ & & $\!\sigma_3\otimes\one\!$ &  &   $\!\sigma_3\otimes\one\otimes\one\!$
\\
\hline
\end{tabular}
\end{equation}


\subsection{Lie algebras associated with the Clifford group}
\label{sec-CGLie}

The operators which are left invariant by the symmetries $\Sigma$ and $\Omega$ form Lie algebras $\mathfrak{g}^d(\Sigma)$ and $\mathfrak{g}^d(\Sigma,\Omega)$ which will be introduced next. The construction is such that $\Gamma_1,\ldots,\Gamma_d$ are all elements of these Lie algebras, and consequently also the Dirac operator defined in \eqref{eq-diracop} (more precisely, it is only in an unbounded version of the Lie algebras). Let $\Hh$ be a Hilbert space with complex conjugation and extend $\Sigma$ and $\Omega$ to $\BM(\Hh)\otimes\CM^{d'\times d'}$ by tensoring with $\one$ where $d^\prime = 2^{\lfloor \frac{d}{2}\rfloor}$. Using the notation $A^{[1]}=A$ and $A^{[-1]}=-A^*$, let us then set, with $\kappa\in\{-1,1\}$ as  in \eqref{tab-normal} and in Proposition~\ref{prop-SigmaUnique},
\begin{align*}
\mathfrak{g}^d(\Sigma)
\;& \; =\;
\left\{
A\in\BM(\Hh)\otimes\CM^{d'\times d'}\;\left|\;
\Sigma^*\,\overline{A}\,\Sigma\,=\,A^{[\kappa]}
\right.
\right\}\;,
\qquad
& d\;\mbox{\rm odd}\;,
\\
\mathfrak{g}^d(\Sigma,\Omega)
\;& \; =\;
\left\{
A\in\BM(\Hh)\otimes\CM^{d'\times d'}\;\left|\;
\Sigma^*\,\overline{A}\,\Sigma\,=\,A^{[\kappa]}\;,
\;\;
\Omega^*\,A\,\Omega=\,-\,A^*
\right.
\right\}\;,
\qquad
& d\;\mbox{\rm even}\;.
\end{align*}
If one works directly with the normal forms of symmetry operators $\Sigma$ and $\Omega$ as given in \eqref{tab-normal}, the dimension $d'$ of the tensored matrix algebra can be reduced to $2$ for $d\not =0\,\mbox{mod}\,4$ and to $4$ for $d =0\,\mbox{mod}\,4$. Let us note that these sets are indeed Lie algebras, namely if $A$ and $B$ are in one of these sets, so is the commutator $[A,B]$. This also justifies the introduction of the notation $A^{[\kappa]}$. Moreover it is possible to replace one of the defining relations in $\mathfrak{g}^d(\Sigma,\Omega)$ for even $d$ by $\widehat{\Sigma}^*\,\overline{A}\,\widehat{\Sigma}\,=\,A^{[-\kappa]}$ so that 
\begin{align*}
\mathfrak{g}^d(\Sigma,\Omega)
 & =\;\left\{
A\in\BM(\Hh)\otimes\CM^{d'\times d'}\;\left|\;
\widehat{\Sigma}^*\,\overline{A}\,\widehat{\Sigma}\,=\,A^{[-\kappa]}\;,
\;\;
\Omega^*\,A\,\Omega=\,-\,A^*
\right.
\right\}
\\
 & =\;\left\{
A\in\BM(\Hh)\otimes\CM^{d'\times d'}\;\left|\;
\Sigma^*\,\overline{A}\,\Sigma\,=\,A^{[\kappa]}\;,
\;\;\widehat{\Sigma}^*\,\overline{A}\,\widehat{\Sigma}\,=\,A^{[-\kappa]}
\right.
\right\}
\;.
\end{align*}
This shows that, indeed, $\Sigma$ and $\widehat{\Sigma}$ can be considered on equal footing for even $d$. Moreover, the in presence of a chiral symmetry the real symmetry can always be chosen in the familiar form $\widetilde{\Sigma}^*\overline{A}\,\widetilde{\Sigma}=A$ if one chooses $\widetilde{\Sigma}$ to be either $\Sigma$ or $\widehat{\Sigma}$. Furthermore, let us point out that  $\mathfrak{g}^{d+8}(\Sigma)\cong\mathfrak{g}^{d}(\Sigma)$ and $\mathfrak{g}^{d+8}(\Sigma,\Omega)\cong\mathfrak{g}^{d}(\Sigma,\Omega)$ in the stabilized regime of infinite dimensional $\Hh$. Moreover, $\mathfrak{g}^d(\Sigma,\Omega)$ is a Lie-subalgebra of $\mathfrak{g}^{d+1}(\Sigma)$ for even $d$:
\begin{equation}
\label{eq-evenSub}
\mathfrak{g}^d(\Sigma,\Omega)
\;=\;
\left\{
A\in\mathfrak{g}^{d+1}(\Sigma)\;\left|\;
\Omega^*\,A\,\Omega\,=\,-\,A^*
\right.
\right\}
\;,
\qquad 
d\;\mbox{\rm even}\;.
\end{equation}
The following result shows that a similar inclusion also holds for odd $d$, provided the correct symmetry operator is used.

\begin{proposi} 
\label{prop-Subgroup}
Set $\widehat{\Omega}=\imath\,\Gamma_{d+1}$. Then
\begin{equation}
\label{eq-oddSub}
\mathfrak{g}^d(\Sigma)
\;\cong\;
\left\{
A\in\mathfrak{g}^{d+1}(\Sigma,\Omega)\;\left|\;
\widehat{\Omega}^*\,A\,\widehat{\Omega}\,=\,-\,A^*
\right.
\right\}
\;,
\qquad 
d\;\mbox{\rm odd}\;.
\end{equation}
\end{proposi}

It is then possible to iterate \eqref{eq-evenSub} and \eqref{eq-oddSub}. We do not write out these formulas, but for this purpose it is clearly advisable to place a dimensional index on $\Sigma^d$, e.t.c.. For the  proof of Proposition~\ref{prop-Subgroup}, a general transformation property of the Lie-algebras under a unitary operator $B$ will be used:
\begin{equation}
\label{eq-LieAlgTrans}
B^*\,\mathfrak{g}^d(\Sigma)\,B
\;=\;
\mathfrak{g}^d(B^t\Sigma B)
\;,
\qquad
B^*\,\mathfrak{g}^d(\Sigma,\Omega)\,B
\;=\;
\mathfrak{g}^d(B^t\Sigma B,B^*\Omega B)
\;.
\end{equation}
If $B$ is a real unitary, then $B^t\Sigma B$ and $B^*\Omega B$ are again symmetry operators in the sense of Definition~\ref{def-SymmetryOp}. For complex $B$, it may be that $\imath \,B^t\Sigma B$ and/or  $\imath\,B^*\Omega B$ are again symmetry operators. Adding the imaginary factor does not modify the algebraic relations though, {\it e.g.} $\mathfrak{g}^d(B^t\Sigma B,B^*\Omega  B)=\mathfrak{g}^d(\imath\,B^t\Sigma B,\imath\,B^*\Omega  B)$. 

\vspace{.2cm}

\noindent {\bf Proof} of Proposition~\ref{prop-Subgroup}. Let $\widehat{\mathfrak{g}}^d$ denote the Lie algebra on the r.h.s. of \eqref{eq-oddSub}. Replacing $\Sigma$ by $\widehat{\Sigma}$, one then has
$$
\widehat{\mathfrak{g}}^d
\,=\,
\left\{
A\in\BM(\Hh)\otimes\CM^{(d+1)'\times (d+1)'}\,\left|\,
\Omega^*A\,\Omega=-A^*\,,\;
\widehat{\Omega}^*A\,\widehat{\Omega}=-A^*\,,\;
\widehat{\Sigma}^*\overline{A}\,\widehat{\Sigma}=A^{[-\kappa]}
\right.
\right\}\,,
$$
where $\kappa=\kappa^{d+1}=-\kappa^d$. Further note that here $\widehat{\Sigma}=\widehat{\Sigma}^{d+1}$, $\Omega=\Omega^{d+1}$ and $\widehat{\Omega}=\widehat{\Omega}^{d+1}$. Now $(d+1)'=2\,d'$ so that $\CM^{(d+1)'\times (d+1)'}=\CM^{d'\times d'}\otimes\CM^{2\times 2}$. A unitary transformation in the last $2\times 2$ component will be done with $B=\one_{d'}\otimes c$ where $c$ is the Cayley transform defined in \eqref{eq-Cayley}. Using the identities \eqref{eq-CayleyShift},
$$
B^*\Omega^{d+1} B\;=\;\Omega^d\otimes\sigma_2\;,
\qquad
B^*\widehat{\Omega}^{d+1} B\;=\;\imath\,\Omega^d\otimes\sigma_1
\;,
\qquad
B^t\widehat{\Sigma}^{d+1} B\;=\;
{\Sigma}^d\otimes\one_2
\;,
$$
where for the last equality the relation \eqref{eq-SigmaHatDef} and $c^t\sigma_1 c=\one_2$ was used. Therefore the relations for $A\in B^*\widehat{\mathfrak{g}}^dB$ are
$$
(\Omega^d\otimes\sigma_2)^*A\,\Omega^d\otimes\sigma_2\;=\;-A^*\;,
\quad
(\Omega^d\otimes\sigma_1)^*A\,\Omega^d\otimes\sigma_1\;=\;-A^*\;,
\quad
({\Sigma}^d\otimes\one_2)^*\overline{A}\, {\Sigma}^d\otimes\one_2=A^{[\kappa^d]}
\;.
$$
Now let us write $A=\binom{a \;\;b}{c\;\;d}$ with $a,b,c,d\in \BM(\Hh)\otimes \CM^{d'\times d'}$. Then the first two relations imply $d=(\Omega^d)^* a^*\Omega^d$ and $b=c=0$, so that the last relation simply becomes $({\Sigma}^d)^*\overline{a}\,{\Sigma}^d=a^{[\kappa_d]}$, which is the defining relation of $\mathfrak{g}^d(\Sigma)$.\
\hfill $\Box$

\vspace{.2cm}

It will follow from the arguments in the next section that in the case of
a finite dimensional Hilbert space $\Hh$ the Lie algebras constructed above provide all the families of Lie algebras of the $8$ classical matrix groups which require a real structure for their definition.

\subsection{Lie groups associated with the Clifford group}
\label{sec-CGLieG}

Associated to Lie algebras one obtains Lie groups by exponentiation $T=e^A$. This is applied to the Lie algebras of Section~\ref{sec-CGLie}. Setting $T^{\{\kappa\}}=e^{A^{[\kappa]}}$, namely $T^{\{1\}}=T$ and $T^{\{-1\}}=(T^*)^{-1}$, one finds
\begin{align*}
\Gg^d(\Sigma)
\;& \; =\;
\left\{
T\in\BM(\Hh)\otimes\CM^{d'\times d'}\;\mbox{\rm invertible}\;\left|\;
\Sigma^*\,\overline{T}\,\Sigma\,=\,T^{\{\kappa\}}
\right.
\right\}\;,
& d\;\mbox{\rm odd}\;,
\\
\Gg^d(\Sigma,\Omega)
\;& \; =\;
\left\{
T\in\BM(\Hh)\otimes\CM^{d'\times d'}\;\mbox{\rm invertible}\;\left|\;
\Sigma^*\,\overline{T}\,\Sigma\,=\,T^{\{\kappa\}}\;,
\;\;
T^*\Omega\,T\,=\,\Omega
\right.
\right\}\;,
& 
d\;\mbox{\rm even}\;.
\end{align*}
Again it is possible to rewrite the groups for even $d$:
\begin{align*}
\Gg^d(\Sigma,\Omega)
\;& \; =\;
\left\{
T\in\BM(\Hh)\otimes\CM^{d'\times d'}\;\mbox{\rm invertible}\;\left|\;
\Sigma^*\,\overline{T}\,\Sigma\,=\,T^{\{\kappa\}}\;,
\;\;
\widehat{\Sigma}^*\,\overline{T}\,\widehat{\Sigma}\,=\,T^{\{-\kappa\}}
\right.
\right\}
\\
\;& \; =\;
\left\{
T\in\BM(\Hh)\otimes\CM^{d'\times d'}\;\mbox{\rm invertible}\;\left|\;
\widehat{\Sigma}^*\,\overline{T}\,\widehat{\Sigma}\,=\,T^{\{-\kappa\}}\;,
\;\;
T^*\Omega\,T\,=\,\Omega
\right.
\right\}
\;.
\end{align*}
%
The main object of the remainder of this section is to show that this construction provides the $8$ real families of classical matrix groups in the case where $\Hh$ is of adequate finite dimension in order to fit with the $n$ below. The table also lists the properties characterizing the elements of the these groups: (even) real, quaternionic (or odd real), (even) symmetric and odd symmetric:

\vspace{.1cm} 

\begin{center}

\begin{tabular}{|c||c|c|c|c|c|c|c|c|}
\hline
$d$ & $1$ & $2$ & $3$ & $4$ & $5$ & $6$ & $7$ & $8$
\\
\hline
\hline
CAZ & AI & CI & C & CII & AII & DIII& D & BDI
\\
\hline
$\Gg^d\cong$ &  $\mbox{\rm Gl}(n,\RM)$ & $\mbox{\rm Sp}(n,\RM)$ & $\mbox{\rm Sp}(n,\CM)$ & $\mbox{\rm Sp}(2n,2n)$ & $\mbox{\rm U}^*(2n)$ & $\mbox{\rm SO}^*(2n)$ & $\mbox{\rm O}(n,\CM)$ & $\mbox{\rm O}(n,n)$   
\\
\hline
 & real & \stackanchor{real}{odd sym} & odd sym & \stackanchor{quat}{odd sym} & quat & \stackanchor{quat}{sym} & sym & \stackanchor{real}{sym} 
\\
\hline
\end{tabular}
\end{center}

\vspace{.1cm} 

\noindent The remaining two complex families (of Weyl's $10$ families of classical matrix groups) are $\mbox{\rm Gl}(n,\CM)$ and $\mbox{\rm U}(n,m)$. In order to verify this table, it is again useful to use the general transformation property of the group under a unitary operator $B$:
$$
B^*\,\Gg^d(\Sigma)\,B
\;=\;
\Gg^d(B^t\Sigma B)
\;,
\qquad
B^*\,\Gg^d(\Sigma,\Omega)\,B
\;=\;
\Gg^d(B^t\Sigma B,B^*\Omega B)
\;.
$$
This corresponds to \eqref{eq-LieAlgTrans} on the level of Lie algebras. First let us choose $B=O$ as in \eqref{tab-normal} so that the symmetry operators are the normal form given in \eqref{tab-normal}. This basis change $O$ will be notationally suppressed in the following. In a second step, it will be helpful to use also a complex $B$, namely the Cayley transform $C=c\otimes\one$. The symmetries and their transformations are then

\vspace{.1cm}

\begin{center}

\begin{tabular}{|c||c|c|c|c|c|c|c|c|c|c|c|c|}
\hline
$d\,\mbox{\rm mod}\,8$ & $1$ & $2$ & $3$ & $4$ & $5$ & $6$ & $7$ & $8$
\\
\hline
\hline
$\kappa=$ & $1$ & $-1$ & $-1$  & $1$  & $1$ & $-1$ & $-1$ & $1$
\\
\hline
$\!\Sigma =\!$ & $\!\sigma_1\otimes\one\!$ &  $\!\imath\sigma_2\otimes\one\!$  & $\!\imath\sigma_2\otimes\one\!$ & $\!\one\otimes\imath\sigma_2\!$  & $\!\imath\sigma_2\otimes\one\!$ & $\!\sigma_1\otimes\one\!$  &  $\!\sigma_1\otimes\one\!$  & $\!\one\otimes\sigma_1\!$ 
\\
$\!\Omega =\!$ &  &  $\!\sigma_3\otimes\one\!$ &  &  $\!\sigma_3\otimes\one\!$ &  & $\!\sigma_3\otimes\one\!$   &  &  $\!\sigma_3\otimes\one\!$
\\
$\!\widehat{\Sigma} =\!$ &  &  $\!\sigma_1\otimes\one\!$  & & $\!\sigma_3\otimes\imath\sigma_2\!$  & & $\!\imath\sigma_2\otimes\one\!$  & &   $\!\sigma_3\otimes\sigma_1\!$ 
\\
\hline
$\!C^t\Sigma C=\!$ & $\!\one\otimes\one\!$ &  $\!-\sigma_2\otimes\one\!$  & $\!-\sigma_2\otimes\one\!$ & $\!\sigma_3\otimes\sigma_2\!$  & $\!-\sigma_2\otimes\one\!$ & $\!\one\otimes\one\!$  & $\!\one\otimes\one\!$ &   $\!\sigma_3\otimes\sigma_1\!$ 
\\
$\!C^*\Omega C=\!$ &  &  $\!\sigma_2\otimes\one\!$ &   & $\!\sigma_2\otimes\one\!$ &  & $\!\sigma_2\otimes\one\!$   &    &$\sigma_2\otimes\one$
\\
$\!C^t\widehat{\Sigma} C=\!$ &  &  $\!\one\otimes\one\!$  & & $\!\sigma_1\otimes\sigma_2\!$  & & $\!-\sigma_2\otimes\one\!$  & &   $\!-\imath\sigma_1\otimes\sigma_1\!$ 
\\
\hline
\end{tabular}
\end{center}

\vspace{.1cm}

\noindent Here the last two rows follow from the identities
$$
c^t\sigma_1 c=\one
\;,
\qquad
c^t\sigma_2 c
\;=\;
\imath\sigma_2
\;,
\qquad
c^t\sigma_3 c\;=\;-\,\imath\sigma_1
\;,
\qquad
c^t c\;=\;\sigma_3
\;,
\qquad
c c^t\;=\;\sigma_1
\;.
$$
The Cayley transforms of the symmetry operators appearing in the lower three rows of the table are not all symmetry operators again because some of them are purely imaginary. This can, however, be readily fixed by multiplying them with the imaginary unit $\imath$. This multiplication does not change the defining relations in  $\Gg^d(C^t\Sigma C,C^*\Omega  C)$. The remarkable fact about the Cayley transform is that it changes the commutation relations between the symmetry operators from commuting to anti-commuting, and vice versa. In particular, one can choose a representation of the $4$ groups in even dimension which only invokes commutating symmetry operators. This is crucial in connection with the commuting symmetries of the Hamiltonian, see Section~\ref{sec-Ham}. Furthermore, the Cayley transform helps to identify the classical groups as claimed above. For example,
\begin{align*}
C^*\,\Gg^1(\Sigma)\,C
\;&  
\; =\;
\left\{
T\in\BM(\Hh)\otimes\CM^{d'\times d'}\;\mbox{\rm invertible}\;\left|\;
\overline{T}\,\,=\,T\;
\right.
\right\}
\;,
\\
C^*\,\Gg^2(\Sigma,\Omega)\,C
\;&  
\; =\;
\left\{
T\in\BM(\Hh)\otimes\CM^{d'\times d'}\;\mbox{\rm invertible}\;\left|\;
\overline{T}\,\,=\,T\;,
\;\;
T^*\imath\sigma_2\,T\,=\,\imath\sigma_2
\right.
\right\}
\;,
\\
C^*\,\Gg^6(\Sigma,\Omega)\,C
\;&  
\; =\;
\left\{
T\in\BM(\Hh)\otimes\CM^{d'\times d'}\;\mbox{\rm invertible}\;\left|\;
(\imath\sigma_2)^*\,\overline{T}\,\imath\sigma_2\,=\,T\;,
\;\;
T^*\imath\sigma_2\,T\,=\,\imath\sigma_2
\right.
\right\}
\\
&  
\; =\;
\left\{
T\in\BM(\Hh)\otimes\CM^{d'\times d'}\;\mbox{\rm invertible}\;\left|\;
(\imath\sigma_2)^*\,\overline{T}\,\imath\sigma_2\,=\,T\;,
\;\;
\overline{T}\,=\,(T^*)^{-1}
\right.
\right\}
\;,
\end{align*}
where $\sigma_2=\sigma_2\otimes\one$. These forms make apparent the entries $\mbox{\rm Gl}(n,\RM)$, $\mbox{\rm Sp}(2n,\RM)$ and $\mbox{\rm SO}^*(2n)$ in the columns $d=1,2,6$ of the above table of classical groups. The others can be checked in the same manner. As a final comment, let us point out that \eqref{eq-evenSub} and \eqref{eq-oddSub} imply that the classical groups form a hierarchy of subgroups (the isomorphisms of Proposition~\ref{prop-Subgroup} are suppressed):
\begin{align*}
\ldots\;
& 
\subset\;\;\,
\mbox{\rm Gl}(n,\RM)
\!\!\!\!\!\! & & \subset\;\;\,
\mbox{\rm Sp}(2n,\RM)
\!\!\!\!\!\! & & \subset\;\;\,
\mbox{\rm Sp}(2n,\CM)
\!\!\!\!\!\! & & \subset\;\;\,
\mbox{\rm Sp}(4n,4n)
\\
&
\subset\;\;\,
\mbox{\rm U}^*(4n)
\!\!\!\!\!\! & & \subset\;\;\,
\mbox{\rm SO}^*(8n)
\!\!\!\!\!\! & & \subset\;\;\,
\mbox{\rm O}(8n,\CM)
\!\!\!\!\!\! & & \subset\;\;\,
\mbox{\rm O}(16n,16n)
\;\subset\;
\mbox{\rm Gl}(16n,\RM)
\;\subset\; \ldots\;
\;.
\end{align*}



\end{document}